\documentclass[%
 reprint,
superscriptaddress,
showpacs,preprintnumbers,
 amsmath,amssymb,
 aps,
 pra,
nobalancelastpage,
letterpaper
]{revtex4-1}

\usepackage[dvipdfmx]{graphicx}% Include figure files
\usepackage{dcolumn}% Align table columns on decimal point
\usepackage{bm}% bold math
\usepackage{amsmath,amsfonts,amssymb,amsthm}
\usepackage{txfonts}
\usepackage{enumerate}
\usepackage{color}
\usepackage{braket}
\usepackage{cases}
\usepackage{comment}
\usepackage{nameref}

\makeatletter

\@addtoreset{equation}{section}
\makeatother

\begin{document}

\preprint{APS/123-QED}

\title{Two-dimensional Schr\"odinger symmetry and three-dimensional breathers and Kelvin-ripple complexes 
as quasi-massive-Nambu-Goldstone modes}
%\title{2D Schr\"odinger symmetry and 3D breathers/Kelvin-ripple complexes \\ 
%as quasi-massive-Nambu-Goldstone modes}
\author{Daisuke A. Takahashi}\email{daisuke.takahashi@keio.jp}
\affiliation{Research and Education Center for Natural Sciences, Keio University, Hiyoshi 4-1-1, Yokohama, Kanagawa 223-8521, Japan}
%\affiliation{Kanagawa Institute of Technology, 1030 Shimo-Ogino, Atsugi, Kanagawa 243-0292, Japan}
%\affiliation{RIKEN Center for Emergent Matter Science (CEMS), Wako, Saitama 351-0198, Japan}
\author{Keisuke Ohashi}
\affiliation{Research and Education Center for Natural Sciences, Keio University, Hiyoshi 4-1-1, Yokohama, Kanagawa 223-8521, Japan}
\author{Toshiaki Fujimori}
\affiliation{Research and Education Center for Natural Sciences, Keio University, Hiyoshi 4-1-1, Yokohama, Kanagawa 223-8521, Japan}
\author{Muneto Nitta}
\affiliation{Research and Education Center for Natural Sciences, Keio University, Hiyoshi 4-1-1, Yokohama, Kanagawa 223-8521, Japan}
\affiliation{Department of Physics, Keio University, Hiyoshi 4-1-1, Yokohama, Kanagawa 223-8521, Japan}

\date{\today}% It is always \today, today,

\begin{abstract}
	Bose-Einstein condensates (BECs) confined in a two-dimensional (2D) harmonic trap are known to possess a hidden 2D Schr\"odinger symmetry, that is, the Schr\"odinger symmetry modified by a trapping potential. Spontaneous breaking of this symmetry gives rise to a breathing motion of the BEC, whose oscillation frequency is robustly determined by the strength of the harmonic trap. In this paper, we demonstrate that the concept of the 2D Schr\"odinger symmetry can be applied to predict the nature of \textit{three dimensional} (3D) collective modes propagating along a condensate confined in an elongated trap. We find three kinds of collective modes whose existence is robustly ensured by the Schr\"odinger symmetry, which are physically interpreted as one breather mode and two Kelvin-ripple complex modes, i.e., composite modes in which the vortex core and the condensate surface oscillate interactively. We provide analytical expressions for the dispersion relations (energy-momentum relation) of these modes using the Bogoliubov theory [D.~A.~Takahashi and M.~Nitta, Ann.~Phys. \textbf{354}, 101 (2015)]. Furthermore, we point out that these modes can be interpreted as ``quasi-massive-Nambu-Goldstone (NG) modes,''  that is, they have the properties of both quasi-NG and massive NG modes: quasi-NG modes appear when a symmetry of a part of a Lagrangian, which is not a symmetry of a full Lagrangian, is spontaneously broken, while massive NG modes appear when a modified symmetry is spontaneously broken.
\end{abstract}

%\pacs{}% PACS, the Physics and Astronomy
\maketitle

\section{Introduction}
\begin{figure}[tb]
	\begin{center}
	\includegraphics{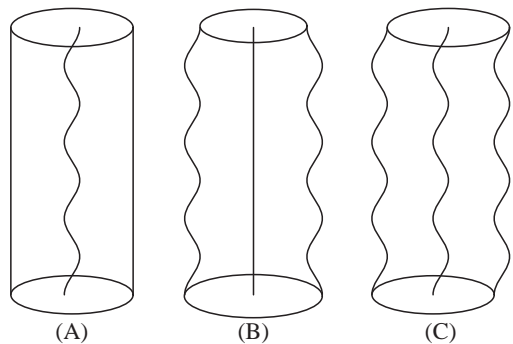}
	\caption{\label{fig:intro} The excited states of an elongated 3D BEC considered in this paper. (A) The pure Kelvin mode, a helical motion of a vortex. (B) The breather mode, an oscillation of the outer condensate. (C) The KR complex mode.  Using the concept of the 2D Schr\"odinger symmetry, we can rigorously find and determine the physical characteristics of excitations (B) and (C). There are one breather corresponding to (B) and two independent KR complex modes corresponding to (C). The pure Kelvin mode (A) cannot be discussed in the framework of the 2D Schr\"odinger symmetry, though we can find it numerically. }
	\end{center}
\end{figure}

	Conformal symmetry plays crucial roles in many branches of modern physics to extract nontrivial consequences which are far from intuition. The non-relativistic $ d $-dimensional Schr\"odinger equation, $ \mathrm{i}\partial_t\psi=-\nabla^2\psi $ with $ \nabla=(\partial_1,\dots,\partial_d) $, has a spacetime scaling symmetry, whose transformation group is called the Schr\"odinger group (or non-relativistic conformal group) denoted by $ \mathrm{Sch}(d) $ (e.g., Refs.~\cite{Hagen:1972pd,Niederer:1972zz,Henkel2003407}).  In particular, the generators of the special Schr\"odinger transformation, the dilatation, and the time-translation form the $ sl(2,\mathbb{R}) $ algebra. \\ 
	\indent The Schr\"odinger symmetry survives under a specific nonlinear generalization. Let us consider the nonlinear Schr\"odinger (NLS) equation,
	\begin{align}
		\mathrm{i}\partial_t\psi=-\nabla^2\psi+2g |\psi|^\alpha \psi,
	\end{align}
	where $ g $ is a nonzero constant. Then, this equation preserves the $ \mathrm{Sch}(d) $ covariance if the power is given by $ \alpha=4, \, 2, $ and $ \frac{4}{3} $ for $ d=1,\, 2, $ and $ 3 $, respectively. The cases  $ (d,\alpha)=(1,4) $ and $ (2,2) $ are important in the physics of ultracold atomic gases, since they represent the dynamics of the Tonks-Girardeau gas in one dimension \cite{PhysRevLett.85.1146,PhysRevA.65.012103} and the Gross-Pitaevskii (GP) equation describing the dynamics of Bose-Einstein condensates (BECs) in two-dimensional (2D) systems. \\
	\indent The above-mentioned $\mathrm{Sch}(d)$ symmetries can be applied, in fact, even in the presence of the harmonic trap,
	\begin{align}
		\mathrm{i}\partial_t\psi=-\nabla^2\psi+2g |\psi|^\alpha \psi+\frac{\omega^2\boldsymbol{r}^2}{4}\psi,
	\end{align}
	which can be experimentally realized by standard techniques in ultracold atomic gases. Due to the existence of the harmonic trap, the original Schr\"odinger symmetry is explicitly broken. However, the harmonic trap term can be eliminated by a certain transformation, so that a modified symmetry exists. Thus, nontrivial time-dependent solutions can be generated from a stationary solution \cite{PhysRevA.55.R853,PhysRevA.65.012103,PhysRevE.64.056602,Ohashi:2017vcy}. The breathing motion of the 2D BEC in a harmonic trap originating from the $ sl(2,\mathbb{R}) $ is theoretically predicted and experimentally observed in Refs.~\cite{PhysRevA.55.R853,PhysRevLett.88.250402}. The Schr\"odinger symmetry in a harmonic trap was discussed in the context of non-relativistic conformal field theory in Ref.\,\cite{Nishida:2007pj, Doroud:2015fsz}. Similar investigations in supersymmetric gauge theories with a harmonic trap ($ \Omega $-background) are found in Ref. \cite{Tong2015}. \\
	\indent The aim of this paper is a different application of 2D Schr\"odinger symmetry --- we apply it to predict the collective modes in  \textit{three-dimensional} (3D) BECs trapped in an elongated trap. 
	Our setup is a harmonic trapping potential in the $xy$-directions and translationally invariance in the $z$-direction, representing a trap elongating in the $z$-direction.
We emphasize that the 3D GP equation itself is \textit{not} covariant under the $\mathrm{Sch}(3)$ operations. Nevertheless,  we find three kinds of collective modes whose existence is robustly guaranteed by the 2D Schr\"odinger symmetry. The physical interpretation of the three modes is: (i) a breather mode, which can be regarded as a generalization of the breathing mode in 2D BEC \cite{PhysRevA.55.R853,Ohashi:2017vcy}, now propagating along the elongated $ z $ direction, and (ii) two Kelvin-ripple (KR) complex modes, i.e., the composite modes consisting of the helical oscillation of the vortex core and the condensate surface. See Fig.~\ref{fig:intro} for their physical picture. Moreover, we derive an analytical expression for the dispersion relations (energy-momentum relations) of these modes, using the Bogoliubov theory approach \cite{Takahashi2015101,PhysRevD.91.025018,PhysRevB.91.184501}. \\ 
	\indent The above-mentioned collective modes are one of the variant concepts of the Nambu-Goldstone (NG) modes emerging in the systems with spontaneous symmetry breaking (SSB). Recently, a general framework of NG modes in non-relativistic systems has been explored \cite{Watanabe:2011ec,Watanabe:2012hr,Hidaka:2012ym}. Our modes have properties of two variants of NG modes: massive NG modes and quasi-NG modes.
	
	When a generalized chemical potential term generated by a symmetry generator is added to the original Hamiltonian or Lagrangian, NG modes are gapped and such modes are called massive NG modes \cite{Nicolis:2012vf,Nicolis:2013sga,PhysRevLett.111.021601}. 
	They are gapful, but their existence is still robustly ensured and the value of the energy gap is determined only by the Lie algebra of the symmetry group.
	Although the authors in Refs.~\cite{Nicolis:2012vf,Nicolis:2013sga,PhysRevLett.111.021601} interpreted that they become massive because of explicit symmetry breaking by the chemical potential term, it is not the case; 
%	If an external field term, which lowers the total Lagrangian symmetry, can be eliminated via some transformation, a gapful mode called the \textit{massive NG mode} appear 
adding chemical potential does not explicitly break the original symmetry but just modifies it. 
This fact was found for the Schr\"odinger symmetry \cite{Ohashi:2017vcy} following the cases of the $O(2,1)$ subgroup \cite{PhysRevA.55.R853,PhysRevA.65.012103} and the Galilean subgroup \cite{PhysRevE.64.056602}. 
Here, a harmonic trapping potential can be introduced as a generalized chemical potential for the special Schr\"odinger symmetry. 
All possible generalized chemical potentials including the rotation generator were introduced in Ref.\,\cite{Ohashi:2017vcy}.
When such modified symmetry is spontaneously broken, there appear massive NG modes. 
Thus, the breathing, harmonic oscillation and cyclotron motions of the 2D trapped BEC can be identified as massive NG modes corresponding to spontaneously broken modified Schr\"odinger symmetry \cite{Ohashi:2017vcy}. 
	
On the other hand, the equation of motion or a part of a Lagrangian sometimes has a symmetry larger than the original Lagrangian.
When such an enhanced symmetry is spontaneously broken, 
there appear \textit{quasi-NG modes} \footnote{
One should not be confused with a similar terminology pseudo-NG modes, which are used for SSB of explicitly broken symmetry, which are massive because of the explicit breaking.}. 
In this case, the ground-state order-parameter space (OPS) is enlarged 
from the original OPS of usual NG modes. Quasi-NG modes become gapped when quantum corrections are taken into account. 
Such situation occurs in pions for chiral symmetry breaking \cite{Weinberg:1972fn}, SSB in supersymmetric field theories 
	%with the potential term possessing a non-compact, complexified symmetry $ G^{\mathbb{C}} $ 
\cite{Kugo:1983ma,Lerche:1983qa,Shore:1984bh,Higashijima:1997ph,Nitta:1998qp,Higashijima:1999ki,Nitta:2014fca,PhysRevD.91.025018}, and condensed matter systems such as superfluid helium 3 \cite{volovik2009universe}, the spin-2 spinor BEC \cite{PhysRevLett.105.230406}, and neutron $^3P_2$ superfluids \cite{Masuda:2015jka}.
Quasi-NG modes which are quite close to the current study appear in a Skyrmion line in magnets \cite{PhysRevD.90.025010}, which has two gapless modes propagating along the line, that is, the dilaton-magnon mode and Kelvin mode. Here, the former corresponds to a spontaneous breaking of the dilational symmetry, which is a symmetry in the 2D section of the Skyrmion line but not a symmetry of full 3D.
	
As mentioned above, we will point out that the collective modes found in this paper satisfy both quasi- and massive conditions --- namely, they are the \textit{``quasi-massive-NG modes.''} \\
\indent Rather surprisingly, a more fundamental low-energy excitation, i.e., the Kelvin mode \cite{Kelvin1880}, cannot be exactly treated by the concept of symmetry in the presence of a trapping potential, and we can find it only numerically. In the case of the Kelvin mode, the vortex core solely oscillates and is not coupled with the condensate-surface oscillation [Fig.~\ref{fig:intro} (A)], in contrast to the KR modes mentioned above. This mode shows the Landau instability (negative dispersion) in the finite-size systems, which is consistent with the case study of the cylindrical trap \cite{Kobayashi01022014,PhysRevB.91.184501}. Studies of the Kelvin modes in ultracold atomic gases in a trapping potential can be found in Refs.~\cite{PhysRevLett.90.100403,PhysRevA.69.043617,PhysRevLett.101.020402}. In the trapless limit, the Kelvin mode becomes an NG mode originating from an SSB of spatial translation, and showing the noninteger dispersion $ \epsilon \sim -k^2 \ln k $ \cite{Pitaevskii1961,Donnelly,PhysRevB.91.184501}. \\
	\indent The organization of this paper is as follows. In Sec.~\ref{sec:def3dproblem}, we introduce the 3D GP and Bogoliubov equations, and clarify the problem discussed in this paper. In Sec.~\ref{sec:2dSchsym}, we summarize the result obtained from the 2D Schr\"odinger symmetry. In particular, we derive the zero-mode solutions of the Bogoliubov equation, which is essential in the Bogoliubov theory approach. Section~\ref{sec:kelvin3D} describes the main results of this paper. We derive the collective excitations originating from the 2D Schr\"odinger symmetry and derive their dispersion relations. We also elucidate their physical picture. In Sec.~\ref{sec:numchk}, we verify our main result by numerical calculations. We also mention the existence of the Kelvin mode [Fig.~\ref{fig:intro}, (A)], which we cannot treat by the concept of the Schr\"odinger symmetry. In Sec.~\ref{sec:QMNG}, we prove that the KR complex and breather modes [Fig.~\ref{fig:intro}, (B),(C)] can be identified as the quasi-massive-NG modes. Section~\ref{sec:summary} is devoted to a summary and outlook.

\section{Definition of the problem}\label{sec:def3dproblem}
	\indent In this section, we clarify the problem which we want to solve in this paper. We start from the 3D GP equation with a harmonic trap in the $x$- and $y$-directions:  
	\begin{align}
		\mathrm{i}\partial_t\psi=-\nabla^2\psi+2g|\psi|^2\psi+\frac{\omega^2(x^2+y^2)}{4}\psi, \label{3DNLS}
	\end{align}
	where $ \nabla^2=\partial_x^2+\partial_y^2+\partial_z^2 $ is the 3D Laplacian, $ g $ is a coupling constant of two-body interaction, and $ \omega>0 $ represents the strength of the harmonic trap. We basically choose a repulsive interaction $ g>0 $ to stabilize vortex solutions, but all results based on the symmetry discussion are valid regardless of the sign of $ g $. \\
	\indent We emphasize that the 3D GP equation (\ref{3DNLS}) itself does \textit{not} have the 3D Schr\"odinger symmetry, i.e., the equation is never covariant under the $ \mathrm{Sch}(3) $ operations. The 3D Schr\"odinger symmetry is retained when the nonlinear term is modified to $ |\psi|^2\psi \to |\psi|^{4/3}\psi $ and the trap is isotropic: $ \frac{\omega^2(x^2+y^2+z^2)}{4} $, but we do not consider such a case in this paper.  \\ 
	\indent Let us introduce the Bogoliubov equation for the Bogoliubov quasiparticles as a linearized small oscillation around a given solution of Eq.\,(\ref{3DNLS}) \cite{Bogoliubov,FETTER197267,DalfovoGiorginiPitaevskiiStringari,PethickSmith}. Substituting $ \psi=\psi+\delta\psi $ in Eq.\,(\ref{3DNLS}) and linearizing the equation w.r.t.  $ \delta\psi $, and writing  $ (u,v)=(\delta\psi,\delta\psi^*) $, we obtain
	\begin{align}
		\mathrm{i}\partial_t \begin{pmatrix}u \\ v \end{pmatrix}=\begin{pmatrix} \hat{H}+4g|\psi|^2 & 2g\psi^2 \\ -2g\psi^{*2} & -\hat{H}-4g|\psi|^2 \end{pmatrix}\begin{pmatrix}u \\ v \end{pmatrix} \label{timedepBogoliubov}
	\end{align}
	with $ \hat{H}=-\nabla^2+\frac{\omega^2(x^2+y^2)}{4} $.  In condensed matter theory, the Bogoliubov equation is frequently used to investigate the nature of low-energy excitations and linear stability for a given stationary state (e.g., Refs.~\cite{FETTER197267,DalfovoGiorginiPitaevskiiStringari,PethickSmith}). \\
	\indent We are interested in a solution of the form 
	\begin{align}
		\psi(t,x,y,z)=\mathrm{e}^{-\mathrm{i}\mu t+\mathrm{i}n\theta}f(r), \label{GPvortexsol}
	\end{align}
	where $ \mu $ is a chemical potential and the cylindrical coordinates are defined by $(x,y,z)=(r\cos\theta,r\sin\theta,z) $. We assume that $ f(r) $ is a non-negative real function and $ n\in \mathbb{Z} $ represents a vortex charge. The differential equation for $ f(r) $ is given by
	\begin{align}
		-f''-\frac{f'}{r}+\left( \frac{n^2}{r^2}-\mu+\frac{\omega^2r^2}{4} \right)f+2gf^3=0. \label{fGP}
	\end{align}
	The particle number per unit length along the $z$-axis, $ N=\int |\psi|^2dxdy  = 2\pi\int rf^2dr $, is a monotonically increasing function of $\mu/\omega$ which vanishes at $\mu=\omega (|n|+1)$.
\\
	\indent For large $gN$ the profile of $ f(r) $ for a vortexless state ($n=0$) is estimated by the Thomas-Fermi (TF) approximation \cite{PethickSmith}, where we ignore the kinetic term. 
	The resultant is
	\begin{align}
		f(r)\simeq f_{\text{TF}}(r):= \sqrt{\frac{4\mu-\omega^2r^2}{8g}}\theta\Big(\frac{2\sqrt{\mu}}{\omega}-r\Big), \label{thomasfermi}
	\end{align}
	which indicates that the position of the condensate surface is estimated as
	\begin{align}
		r\simeq r_{\mathrm{TF}}:=\frac{2\sqrt{\mu}}{\omega}. \label{surfaceposition}
	\end{align}
	We can numerically check that this surface position $ r_{\mathrm{TF}} $ is also valid for vortex states with small $ n $'s. Therefore, we can use $ r_{\text{TF}} $ as an effective system size. \\
	\indent The stationary Bogoliubov equation for the eigenenergy $ \epsilon $  is obtained by setting
	\begin{align}
		\begin{pmatrix}u(t,x,y,z) \\ v(t,x,y,z) \end{pmatrix}=\mathrm{e}^{\mathrm{i}(-\mu t+n\theta)\sigma_3}\mathrm{e}^{\mathrm{i}(-\epsilon t+m\theta+kz)} \begin{pmatrix}u(r) \\ v(r) \end{pmatrix}, \label{statbogoderive}
	\end{align}
	and the resultant equation is
	\begin{align}
		\epsilon\begin{pmatrix} u \\ v \end{pmatrix}&=(\hat{L}_m+\sigma_3k^2)\begin{pmatrix} u \\ v \end{pmatrix} \label{statBogoliubov}, 
	\end{align}
	with
	\begin{align}
		\hat{L}_m&=\begin{pmatrix} \hat{H}_{m+n}-\mu+4gf^2 & 2gf^2 \\ -2gf^2 & -\hat{H}_{m-n}+\mu-4gf^2 \end{pmatrix},\\
		 \sigma_3&=\begin{pmatrix} 1 & 0 \\ 0 &-1 \end{pmatrix}, \\
		 \hat{H}_{m\pm n}&=-\partial_r^2-\frac{1}{r}\partial_r+\frac{(m\pm n)^2}{r^2}+\frac{\omega^2r^2}{4}. \label{statBogoliubov4}
	\end{align}
	 $ \epsilon $ is an eigenenergy of the Bogoliubov quasiparticles, and  $ m \in\mathbb{Z},\ k \in \mathbb{R} $ are quantum numbers labeling eigenstates, indicating the quantized angular momentums and wavenumbers in the $z$-direction, respectively. \\ 
	\indent The stationary Bogoliubov equation (\ref{statBogoliubov}) always provides a pair of eigenstates $ w=(u,v)^T $ with numbers $ (\epsilon,k,m) $ and  $ \sigma_1 w^*=(v^*,u^*)^T $ with $ (-\epsilon^*,-k,-m) $. \\
\indent Most eigenstates are determined only numerically. However, as we will see later, several important low-energy excitations can be identified only by symmetry considerations. Besides, we can calculate their dispersion relations (energy-momentum relation) $ \epsilon=\epsilon(k) $ using the exact eigenfunctions found from the symmetry. The aim of this paper is therefore phrased as follows: \textit{Find all collective modes whose existence is robustly ensured by the symmetry, and determine their energy gaps and dispersion relations.} \\
	 \indent We solve the above-mentioned problem by the Bogoliubov theory approach \cite{Takahashi2015101,PhysRevD.91.025018,PhysRevB.91.184501}. The method consists of two procedures: (i) First, we construct a one-parameter family of the solutions to the GP equation, and by differentiating it, we get a zero-mode solution to the Bogoliubov equation. (ii)  Next, regarding the $ \sigma_3 k^2 $ term as a perturbation term, we solve the finite-wavenumber problem by perturbation theory, and obtain the expansion of the dispersion relation $ \epsilon=\epsilon_0+\epsilon_2 k^2+\dotsb. $ (For the linear dispersion relation, the perturbative expansion needs a modification \cite{Takahashi2015101,PhysRevD.91.025018}.  See the example of the Bogoliubov sound wave in Subsec.~\ref{sec:dispersion}.) \\
	 \indent Equation~(\ref{statBogoliubov}) reduces to the 2D equation if $ k=0 $, and hence the 2D result coming from the 2D Schr\"odinger symmetry is partially applicable. We see this in the next section. 

\section{Consequences of 2D Schr\"odinger symmetry}\label{sec:2dSchsym}
In this section, we summarize the 2D Schr\"odinger symmetry of the harmonically-trapped 2D NLS systems, and derive the SSB-originated zero-mode solutions for the Bogoliubov equation. \\
\indent Henceforth, the algebra of the 2D Schr\"odinger group $ \mathrm{Sch}(2) $ is denoted by $ \mathrm{sch}(2) $. 
\subsection{Schr\"odinger symmetry in the 2D NLS systems with a harmonic trap}\label{subsec:2dSchsymGP}
	\indent Let us consider the 2D NLS (or GP) equation with and without a harmonic trap:
	\begin{align}
		\mathrm{i}\psi_t&=-\psi_{xx}-\psi_{yy}+2g|\psi|^2\psi, \label{free2DNLS} \\
		\mathrm{i}\psi_t&=-\psi_{xx}-\psi_{yy}+2g|\psi|^2\psi+\frac{\omega^2(x^2+y^2)}{4}\psi. \label{trapped2DNLS}
	\end{align}
	\indent Following Ref. \cite{PhysRevA.65.012103}, we consider the following function-to-function map $ \mathcal{T} $:  
	\begin{align}
		\mathcal{T}[\psi](t,x,y)&:=\sqrt{\dot{\tau}(t)}\exp\left( \frac{-\mathrm{i}\ddot{\tau}(t)(x^2+y^2)}{8\dot{\tau}(t)} \right) \times \nonumber \\
		&\qquad\qquad\psi\Big(\tau(t),x\sqrt{\dot{\tau}(t)},y\sqrt{\dot{\tau}(t)}\Big), \label{ghosh2001}
	\end{align}
	where $ \tau(t) $ is a non-decreasing function. Then we can prove the following: if $ \psi $ satisfies Eq.\,(\ref{free2DNLS}),  $ \mathcal{T}[\psi] $ satisfies Eq.\,(\ref{trapped2DNLS}) with the harmonic trap replaced by the time-dependent one $ \frac{\omega(t)^2(x^2+y^2)}{4} $, where $ \omega(t)^2=\frac{1}{2}S\tau(t) $, and $ S\tau(t):= \frac{\dddot{\tau}(t)}{\dot{\tau}(t)}-\frac{3}{2}\left( \frac{\ddot{\tau}(t)}{\dot{\tau}(t)} \right)^2 $ is the Schwarzian derivative.\\
	\indent If  $ \tau(t) $ is chosen to be a fractional linear transformation $ \tau(t)=\frac{at+b}{ct+d},\ ad-bc=1 $, then the Schwarzian derivative vanishes and the map (\ref{ghosh2001}) produces a new solution of Eq.\,(\ref{free2DNLS}), and generators of such transformations forms the algebra $ sl(2,\mathbb{R}) $, which is a subalgebra of $ \mathrm{sch}(2) $.  More general functions $ \tau(t) $ induce extended Schr\"odinger transformations (e.g., Ref.~\cite{Henkel2003407}), whose algebra is also called the Schr\"odinger-Virasoro algebra. The 2D NLS equation is \textit{not} invariant under transformations with general $ \tau(t) $'s and hence the equation changes. The map (\ref{ghosh2001}) generally gives a time-dependent harmonic trap. The \textit{time-independent} trap is obtained by \cite{PhysRevA.65.012103}
	\begin{align}
		\mathcal{T}_{\text{trap}}[\psi](t,x,y)&:=\frac{\exp\left( -\frac{\mathrm{i}}{4}(x^2+y^2)\omega \tan \omega t \right)}{\cos \omega t}\times \nonumber \\
		&\qquad\quad \psi\left( \frac{\tan \omega t}{\omega},\frac{x}{\cos\omega t}, \frac{y}{\cos\omega t} \right),
	\end{align}
	which is realized by the choice $ \tau(t)=\frac{\tan\omega t}{\omega} $. If $ \psi $ satisfies the free 2D NLS equation (\ref{free2DNLS}), then $ \mathcal{T}_{\text{trap}}[\psi] $ satisfies the trapped 2D NLS equation (\ref{trapped2DNLS}). We further define 
	\begin{align}
		&\mathcal{T}_{\text{release}}[\psi](t,x,y):=\nonumber \\
		&\frac{\exp\left( \frac{\mathrm{i}\omega^2t(x^2+y^2)}{4(1+\omega^2t^2)} \right)}{\sqrt{1+\omega^2t^2}} \psi\left( \frac{\tan^{-1}\omega t}{\omega}, \frac{x}{\sqrt{1+\omega^2t^2}}, \frac{y}{\sqrt{1+\omega^2t^2}} \right),
	\end{align}
	which is realized by  $ \tau(t)=\frac{\tan^{-1}\omega t}{\omega} $. If $ \psi $ satisfies Eq.\,(\ref{trapped2DNLS}), then $ \mathcal{T}_{\text{release}}[\psi] $ satisfies Eq.\,(\ref{free2DNLS}). The latter is an inverse of the former, i.e., $ \mathcal{T}_{\text{release}}=\mathcal{T}_{\text{trap}}^{-1} $, and the relation
	\begin{align}
		\mathcal{T}_{\text{trap}}\circ\mathcal{T}_{\text{release}}[\psi]=\mathcal{T}_{\text{release}}\circ\mathcal{T}_{\text{trap}}[\psi]=\psi
	\end{align}
	holds. \\
	\indent Next, let us take $ \hat{Q} \in \mathrm{sch}(2) $, and let us define $ \mathcal{T}_{\text{Sch(2)}}(\alpha):=\mathrm{e}^{\mathrm{i}\alpha \hat{Q}} $. If  $ \psi $ is a solution to the 2D NLS equation (\ref{free2DNLS}), then $ \mathcal{T}_{\text{Sch(2)}}(\alpha)[\psi]=\mathrm{e}^{\mathrm{i}\alpha\hat{Q}}\psi $ is also a solution to the same equation.\\
	\indent Finally, we define
	\begin{align}
		\phi(t,x,y,\alpha)=\mathcal{T}_{\text{trap}}\circ \mathcal{T}_{\text{Sch(2)}}(\alpha)\circ\mathcal{T}_{\text{release}}[\psi](t,x,y). \label{oneparafamily}
	\end{align} 
	Following the above-mentioned properties, if $ \psi(t,x,y) $ satisfies the trapped 2D NLS equation (\ref{trapped2DNLS}), then $ \phi(t,x,y,\alpha) $ also satisfies the same equation. Thus, we get a family of solutions for Eq.\,(\ref{trapped2DNLS}) parametrized by a parameter $ \alpha $.   
	Taking  infinitesimal $\alpha$,  a set of operators  $\{ 
	  \mathcal{T}_{\text{trap}} \circ \hat Q \circ \mathcal{T}_{\text{release}} |\hat Q \in {\rm sch}(2)  
	\} $ turns out to give the generators of the modified symmetry of the model (\ref{3DNLS}), which has been directly discussed in \cite{Ohashi:2017vcy}. 

\subsection{SSB-originated zero mode solutions for the Bogoliubov equation}\label{sec:ssbzero}
	\indent Using the one-parameter family of solutions (\ref{oneparafamily}) for Eq.\,(\ref{trapped2DNLS}), we can now apply the formulation of NG modes based on the Bogoliubov theory \cite{Takahashi2015101,PhysRevB.91.184501,PhysRevD.91.025018}. The zero modes based on the symmetry and parameter derivatives have been discussed in the earlier paper \cite{Takahashi20121589}. \\
	\indent Let us introduce the Bogoliubov equation as a linearization of the GP equation (\ref{trapped2DNLS}): 
	\begin{align}
		\mathrm{i}\partial_t \begin{pmatrix}u \\ v \end{pmatrix}=\begin{pmatrix} \hat{H}+4g|\psi|^2 & 2g\psi^2 \\ -2g\psi^{*2} & -\hat{H}-4g|\psi|^2 \end{pmatrix}\begin{pmatrix}u \\ v \end{pmatrix}, \label{tdBogoliubov2d}
	\end{align}
	with $ \hat{H}=-\partial_x^2-\partial_y^2+\frac{\omega^2(x^2+y^2)}{4} $.
	 Now let us derive the SSB-originated zero-mode solutions \cite{Takahashi2015101}. If we differentiate Eq.\,(\ref{trapped2DNLS}) with  $ \psi $ replaced by $ \phi $ in Eq. (\ref{oneparafamily}) by $ \alpha $ and set $ \alpha=0 $ after differentiation, we get a particular solution for the time-dependent Bogoliubov equation (\ref{tdBogoliubov2d}): 
	\begin{align}
		w=\begin{pmatrix} [\partial_\alpha\phi]_{\alpha=0} \\ [\partial_\alpha\phi]^*_{\alpha=0}  \end{pmatrix}=\begin{pmatrix} \mathcal{T}_{\text{trap}}\circ(\mathrm{i}\hat{Q})\circ\mathcal{T}_{\text{trap}}^{-1}[\psi] \\ \left(\mathcal{T}_{\text{trap}}\circ(\mathrm{i}\hat{Q})\circ\mathcal{T}_{\text{trap}}^{-1}[\psi]\right)^* \end{pmatrix}. \label{zeromodew}
	\end{align}
	This expression provides a formula analogous to Eq.(G.8) in the Appendix G of Ref.~\cite{Takahashi2015101}, where massive NG modes are formulated in terms of the Bogoliubov theory. Choosing $ \hat{Q} $ from  various elements of $ \mathrm{sch}(2) $, we obtain several zero-mode solutions. (Note that in the case of massive NG modes, we get the finite-energy eigenstates of the Bogoliubov equation, but we still keep to use the term ``zero-mode solutions'' for brevity.) Since $ \mathrm{sch}(2) $ is nine-dimensional, we obtain nine solutions of the form (\ref{zeromodew}), unless $ \psi $ preserves some symmetry. The list of the solutions is shown below. We also attach expressions in the trap-free limit $ \omega \to 0 $, which obviously corresponds to $ \mathcal{T}_{\text{trap}}=\text{id} $ and hence given by $ w=(\mathrm{i}\hat{Q}\psi,-\mathrm{i}\hat{Q}^*\psi^*)^T $.
	\begin{enumerate}[(i)]\setlength{\itemsep}{-.5ex}
		\item the $x$-translation $\hat{Q}=-\mathrm{i}\partial_x$:
				\begin{align}
					\mathrm{e}^{\mathrm{i}\alpha \hat{Q}}\psi(t,x,y)&=\psi(t,x+\alpha,y). \\
					w_{x\text{-trans}}&=\frac{\mathrm{i}x\omega\sin\omega t}{2}\begin{pmatrix} \psi \\ -\psi^* \end{pmatrix}+\cos\omega t\begin{pmatrix} \psi_x \\ \psi_x^* \end{pmatrix} \label{zeroxtrans} \\ 
					&\overset{\omega \to 0}{\longrightarrow} \begin{pmatrix} \psi_x \\ \psi_x^* \end{pmatrix}. 
				\end{align}
		\item the $y$-translation $\hat{Q}=-\mathrm{i}\partial_y $:
				\begin{align}
					\mathrm{e}^{\mathrm{i}\alpha \hat{Q}}\psi(t,x,y)&=\psi(t,x,y+\alpha). \\
					w_{y\text{-trans}}&=\frac{\mathrm{i}y\omega\sin\omega t}{2}\begin{pmatrix} \psi \\ -\psi^* \end{pmatrix}+\cos\omega t\begin{pmatrix} \psi_y \\ \psi_y^* \end{pmatrix} \\
					& \overset{\omega \to 0}{\longrightarrow} \begin{pmatrix} \psi_y \\ \psi_y^* \end{pmatrix}.
				\end{align}
		\item the $z$-axis rotation $\hat{Q}=-\mathrm{i}(x\partial_y-y\partial_x) $:
				\begin{align}
					\mathrm{e}^{\mathrm{i}\alpha \hat{Q}}\psi(t,x,y)&=\psi(t,x\cos\alpha-y\sin\alpha,x\sin\alpha+y\cos\alpha).\\
					w_{z\text{-rot}}&=\begin{pmatrix} x\psi_y-y\psi_x \\ x\psi^*_y-y\psi^*_x \end{pmatrix}. \quad \text{($\omega$-independent)} \label{zrotzeromode}
				\end{align}
		\item phase multiplication $ \hat{Q}=1  $: 
				\begin{align}
					\mathrm{e}^{\mathrm{i}\alpha \hat{Q}}\psi(t,x,y)&=\mathrm{e}^{\mathrm{i}\alpha}\psi(t,x,y). \\
					w_{\text{phase}}&=\begin{pmatrix} \mathrm{i}\psi \\ -\mathrm{i}\psi^* \end{pmatrix}. \quad \text{($\omega$-independent)}
				\end{align}
		\item the $x$-Galilei transformation $ \hat{Q}=-x-2\mathrm{i}t\partial_x $: 
				\begin{align}
					\mathrm{e}^{\mathrm{i}\alpha \hat{Q}}\psi(t,x,y)&=\mathrm{e}^{-\mathrm{i}(\alpha x+\alpha^2t)}\psi(t,x+2\alpha t,y). \\
					w_{x\text{-Gal}}&=-\mathrm{i}x\cos\omega t\begin{pmatrix} \psi \\ -\psi^* \end{pmatrix}+\frac{2\sin\omega t}{\omega}\begin{pmatrix} \psi_x \\ \psi_x^* \end{pmatrix} \\
					& \overset{\omega \to 0}{\longrightarrow} -\mathrm{i}x \begin{pmatrix}\psi \\ -\psi^* \end{pmatrix}+2t\begin{pmatrix} \psi_x \\ \psi_x^* \end{pmatrix}. \label{xgalzeroomegazero}
				\end{align}
		\item the $y$-Galilei transformation $ \hat{Q}=-y-2\mathrm{i}t\partial_y$: 
				\begin{align}
					\mathrm{e}^{\mathrm{i}\alpha \hat{Q}}\psi(t,x,y)&=\mathrm{e}^{-\mathrm{i}(\alpha y+\alpha^2t)}\psi(t,x,y+2\alpha t). \\
					w_{y\text{-Gal}}&=-\mathrm{i}y\cos\omega t\begin{pmatrix} \psi \\ -\psi^* \end{pmatrix}+\frac{2\sin\omega t}{\omega}\begin{pmatrix} \psi_y \\ \psi_y^* \end{pmatrix} \\
					&\overset{\omega \to 0}{\longrightarrow} -\mathrm{i}y \begin{pmatrix}\psi \\ -\psi^* \end{pmatrix}+2t\begin{pmatrix} \psi_y \\ \psi_y^* \end{pmatrix}. \label{ygalzeroomegazero}
				\end{align}
		\item the $t$-translation $ \hat{Q}=-\mathrm{i}\partial_t $: 
				\begin{align}
					\mathrm{e}^{\mathrm{i}\alpha\hat{Q}}\psi(t,x,y)&=\psi(t+\alpha,x,y). \\
					w_{t\text{-trans}}&=\cos^2\omega t\begin{pmatrix} \psi_t \\ \psi^*_t \end{pmatrix}-\frac{\omega\sin2\omega t}{2}\begin{pmatrix} \psi+x\psi_x+y\psi_y \\ \psi^*+x\psi^*_x+y\psi^*_y \end{pmatrix}\nonumber \\
					&\quad\qquad+\frac{\mathrm{i}(x^2+y^2)\omega^2\cos2\omega t}{4}\begin{pmatrix}\psi \\ -\psi^*\end{pmatrix} \\
					& \overset{\omega \to 0}{\longrightarrow} \begin{pmatrix} \psi_t \\ \psi^*_t \end{pmatrix}.
				\end{align}
		\item the dilatation  $ \hat{Q}=-\mathrm{i}(1+x\partial_x+y\partial_y+2t\partial_t) $: 
				\begin{align}
					\mathrm{e}^{\mathrm{i}\alpha\hat{Q}}\psi(t,x,y)&=\mathrm{e}^\alpha\psi(\mathrm{e}^{2\alpha}t,\mathrm{e}^\alpha x,\mathrm{e}^\alpha y). \\
					w_{\text{dila}}&=\cos2\omega t\begin{pmatrix} \psi+x\psi_x+y\psi_y \\ \psi^*+x\psi^*_x+y\psi^*_y \end{pmatrix}+\frac{\sin2\omega t}{\omega}\begin{pmatrix} \psi_t \\ \psi_t^* \end{pmatrix}\nonumber \\
					&\qquad+\frac{\mathrm{i}(x^2+y^2)\omega\sin2\omega t}{2}\begin{pmatrix}\psi \\ -\psi^*\end{pmatrix} \\
					& \overset{\omega \to 0}{\longrightarrow} \begin{pmatrix} \psi+x\psi_x+y\psi_y \\ \psi^*+x\psi^*_x+y\psi^*_y \end{pmatrix}+2t\begin{pmatrix} \psi_t \\ \psi_t^* \end{pmatrix}. \label{dilazeroomegazero}
				\end{align}
		\item the special Schr\"odinger transformation \\ $\hat{Q}=\mathrm{i}t(1+x\partial_x+y\partial_y+t\partial_t)+\frac{x^2+y^2}{4} $: 
				\begin{align}
					&\mathrm{e}^{\mathrm{i}\alpha \hat{Q}}\psi(t,x,y)=\tfrac{\exp(\frac{\mathrm{i}\alpha(x^2+y^2)}{4(1+\alpha t)})}{1+\alpha t}\psi(\tfrac{t}{1+\alpha t},\tfrac{x}{1+\alpha t},\tfrac{y}{1+\alpha t}). \\ 
					&w_{\text{special}}=-\frac{\sin^2\omega t}{\omega^2}\begin{pmatrix} \psi_t \\ \psi^*_t \end{pmatrix}-\frac{\sin2\omega t}{2\omega}\begin{pmatrix} \psi+x\psi_x+y\psi_y \\ \psi^*+x\psi^*_x+y\psi^*_y \end{pmatrix}\nonumber \\
					&\qquad\qquad\qquad\qquad+\frac{\mathrm{i}(x^2+y^2)\cos2\omega t}{4}\begin{pmatrix}\psi \\ -\psi^* \end{pmatrix}\label{zerospecial}\\
					&\overset{\omega \to 0}{\longrightarrow} -t^2\begin{pmatrix} \psi_t \\ \psi^*_t \end{pmatrix}-t\begin{pmatrix} \psi+x\psi_x+y\psi_y \\ \psi^*+x\psi^*_x+y\psi^*_y \end{pmatrix}+\frac{\mathrm{i}(x^2+y^2)}{4}\begin{pmatrix}\psi \\ -\psi^* \end{pmatrix}.  \label{speschzeroomegazero}
				\end{align}
	\end{enumerate}
	Note that the ordinary $ t $-translation zero mode is reproduced by $ w_{t\text{-trans}}-\omega^2 w_{\text{special}}=(\psi_t,\psi_t^*)^T $.\\
	\indent  We thus get nine solutions both for free ($ \omega=0 $) and trapped ($ \omega \ne 0 $) systems. However, if we consider a one-vortex solution in the trap-free ($\omega=0$) system, where $ \psi $ tends to a constant value in the spatial infinity $ |\psi|^2 \to \rho_\infty $, the zero-mode solutions created from the rotation (\ref{zrotzeromode}), the Galilei transformation (\ref{xgalzeroomegazero}) and (\ref{ygalzeroomegazero}), the dilatation (\ref{dilazeroomegazero}), and the special Schr\"odinger transformation (\ref{speschzeroomegazero}) contain polynomial coefficients of  $ x,y, $ and $ t $. Hence, they are excluded from physical consideration due to an unphysical divergence at spatial and/or temporal infinities.  Thus, for the stationary solution of the trap-free system, we only get three physical solutions: $ w_{x\text{-trans}},\ w_{y\text{-trans}},  $ and $ w_{\text{phase}} $.  This analysis provides a more simplified solution for the redundancy problem of NG modes \cite{Watanabe:2013iia}. \\
	\indent On the other hand, in the trapped system ($\omega \ne0$),  $ \psi $ decays in a Gaussian fashion, and the polynomial coefficient of $ x,y $ causes no problem in convergence. The $ t $-dependence also appears in the form of $ \cos\omega t $ or $ \sin \omega t $, which remains finite for all time. Thus, all zero-mode solutions created from all elements in  $ \mathrm{sch}(2) $ can contribute as independent  physical modes. This situation is highly contrasted with the trap-free system.

\section{Collective excitations propagating along elongated BECs in 3D}\label{sec:kelvin3D}
\subsection{Seed zero  modes}
	 \indent Equation~(\ref{statBogoliubov}) reduces to the 2D equation if $ k=0 $, and thus the SSB-originated zero-mode solutions derived in Sec.~\ref{sec:ssbzero} can be applied.  Though the solutions (\ref{zeroxtrans})-(\ref{zerospecial}) are time-dependent, we can make a stationary eigenstate for the stationary Bogoliubov equation (\ref{statBogoliubov}) by taking their linear combination. 
%	 They are summarized as follows.
	The eigenstates with their eigenenergies $ \epsilon $, quantized angular momentums $ m $, and wavenumbers $ k $ [introduced in Eqs. (\ref{statbogoderive})-(\ref{statBogoliubov4})] are summarized as follows:  
	\begin{enumerate}[(a)]\setlength{\itemsep}{-1ex}
		\item Seed of the Bogoliubov sound wave, $(\epsilon,k,m)=(0,0,0)$:
		\begin{align}
			w_{\text{phonon}}&=\begin{pmatrix} f \\ -f \end{pmatrix}. \label{zerophonon} %,\\
		\end{align}
		This is made from $ w_{\text{phase}} $ or $ w_{z\text{-rot}} $ or $ w_{t\text{-trans}}-\omega^2 w_{\text{special}} $.
		\item Seed of the first KR complex, $ (\epsilon,k,m)=(\omega,0,-1) $: 
		\begin{align}
			w_{\text{KR1}}&=f'\begin{pmatrix} 1 \\ 1 \end{pmatrix}+\frac{(2n-\omega r^2)f}{2r}\begin{pmatrix} 1 \\ -1 \end{pmatrix}. \label{zerokb1} %,\\
		\end{align}
		This is made by $ w_{x\text{-trans}}-\mathrm{i}w_{y\text{-trans}}-\frac{\mathrm{i}\omega}{2}(w_{x\text{-Gal}}-\mathrm{i}w_{y\text{-Gal}}) $.
		\item Seed of the second KR complex, $(\epsilon,k,m)=(\omega,0,1)$: 
		\begin{align}
			w_{\text{KR2}}&=f'\begin{pmatrix} 1 \\ 1 \end{pmatrix}-\frac{(2n+\omega r^2)f}{2r}\begin{pmatrix} 1 \\ -1 \end{pmatrix}. \label{zerokb2}
		\end{align} 
		This is made by $ w_{x\text{-trans}}+\mathrm{i}w_{y\text{-trans}}-\frac{\mathrm{i}\omega}{2}(w_{x\text{-Gal}}+\mathrm{i}w_{y\text{-Gal}}) $.
		\item Seed of the breather mode, $ (\epsilon,k,m)=(2\omega,0,0) $: 
		\begin{align}
			w_{\text{breather}}&=\omega h' \begin{pmatrix}1 \\ 1\end{pmatrix}+\frac{(2\mu-\omega^2r^2)h}{2r}\begin{pmatrix}1 \\ -1\end{pmatrix}, \quad h(r):=rf(r). \label{zerobreather}
		\end{align}
		This is made by $\omega w_{\text{dila}}+\mathrm{i}(w_{t\text{-trans}}+\omega^2w_{\text{special}})$.
	\end{enumerate}

	\indent In addition to the above four solutions (\ref{zerophonon})-(\ref{zerobreather}), by the symmetry of the Bogoliubov equation, we have conjugate solutions  $ \sigma_1 w_{\text{KR1}},\ \sigma_1 w_{\text{KR2}}, $ and $ \sigma_1w_{\text{breather}} $, whose eigenenergies and quantum numbers are given by $ (\epsilon,k,m)=(-\omega,0,1),\ (-\omega,0,-1), $ and $ (-2\omega,0,0) $, respectively. Thus, we have seven solutions. 
	Since $ w_{\text{phase}},\ w_{z\text{-rot}}, $ and $ w_{t\text{-trans}}-\omega^2 w_{\text{special}} $ are degenerate when $ \psi $ has the form (\ref{GPvortexsol}), all possible solutions obtainable from the nine solutions (\ref{zeroxtrans})-(\ref{zerospecial}) are exhausted by these seven solutions. \\
	\indent A physical picture of the above modes can be understood by considering the density and phase oscillations as follows. 
	If an eigenstate of the Bogoliubov equation $ \alpha w=\alpha (u,v)^T $ is excited, where $ \alpha>0 $ is a small factor justifying the linearization in the derivation of the Bogoliubov equation, the small oscillation of the condensate is given by $ \psi+\delta\psi=\psi+\alpha (u+v^*)  $. In particular, when the solution is given by the form (\ref{GPvortexsol}) and (\ref{statbogoderive}), we obtain
	\begin{align}
		&\psi+\delta\psi=\mathrm{e}^{\mathrm{i}(-\mu t+n\theta)}\times \nonumber \\
		& \left[ f+\alpha\frac{u+v}{2}\cos(kz-\epsilon t+m\theta)+\mathrm{i}\alpha\frac{u-v}{2}\sin(kz-\epsilon t+m\theta) \right]. \label{densityandphase}
	\end{align}
	From this, we can show that the density and phase fluctuations are expressed by 
	\begin{align}
		|\psi+\delta\psi|^2&=f(r)^2+\alpha \delta \rho(r) \cos(kz-\epsilon t+m\theta), \label{densityfluc} \\
		\arg (\psi+\delta\psi)&=-\mu t+n\theta+\alpha \delta\varphi(r) \sin(kz-\epsilon t+m\theta)
	\end{align}
	with
	\begin{align}
		\delta\rho(r):=[u(r)+v(r)]f(r),\quad \delta\varphi(r):=\frac{u(r)-v(r)}{2f(r)}. \label{densityphasefluct}
	\end{align}
	Using the expressions~(\ref{densityandphase})-(\ref{densityphasefluct}), we can find the physical interpretation for a given eigenstate. For example, $ w_{\text{phonon}} $ only has a phase fluctuation, consistent with the fact that this mode originates from the SSB of the $U(1)$ phase transformation. The details of other modes will be discussed in the next subsection. % Subsec.~\ref{sec:dispersion}.

\subsection{Dispersion relation}\label{sec:dispersion}
	We now determine the dispersion relations $ \epsilon(k) $. We solve Eq.\,(\ref{statBogoliubov}) for finite  $ k^2 $  perturbatively, regarding $ \hat{L}_m $ and $ \sigma_3k^2 $ as an unperturbed term and a perturbation term, respectively  \cite{Takahashi2015101,PhysRevB.91.184501,PhysRevD.91.025018}. 
	If we have an eigenstate $ w_0=(u_0,v_0)^T $ with eigenvalue $ \epsilon_0 $, the leading-order correction of the eigenvalue is given by:  
	\begin{align}
		\epsilon(k)=\epsilon_0+\frac{(w_0,\sigma_3w_0)_\sigma}{(w_0,w_0)_\sigma}k^2+O(k^4), \label{dispersiongen}
	\end{align}
	where the $ \sigma $-inner product for $ w_1=(u_1,v_1)^T $ and $ w_2=(u_2,v_2)^T $ is defined by
	\begin{align}
		(w_1,w_2)_\sigma=\int_0^{2\pi}\mathrm{d}\theta\int_0^\infty r\mathrm{d}r (u_1^*u_2-v_1^*v_2).
	\end{align} 
	This product is invariant under the Bogoliubov transformations and used to formulate the perturbation theory \cite{Takahashi2015101}. 
%	A comparison between bosonic and fermionic Bogoliubov transformations is given in Sec.~V of Ref.~\cite{PhysRevB.93.024512}.
	 $ w_0 $ is said to have finite norm if $ (w_0,w_0)_\sigma \ne0 $, and the perturbation theory for finite-norm eigenfunctions is almost the same as that for the well-known hermitian operators. 
	Among the zero-mode solutions (\ref{zerophonon})-(\ref{zerobreather}), only $ w_{\text{phonon}} $ has zero norm, while $ w_{\text{KR1}},\ w_{\text{KR2}}, $ and $ w_{\text{breather}} $ have finite and positive norm.
	Therefore, for the latter three modes, the coefficient of $ k^2 $ is calculated by
	\begin{align}
		\frac{(w_0,\sigma_3w_0)_\sigma}{(w_0,w_0)_\sigma}=\frac{\int r\mathrm{d}r(|u_0|^2+|v_0|^2)}{\int r\mathrm{d}r(|u_0|^2-|v_0|^2)}. \label{dispersiongencoef}
	\end{align}
	In the following, we provide the dispersion relations for these modes.
\subsubsection{$w_{\mathrm{KR1}}$ and  $ w_{\mathrm{KR2}} $ --- the KR complex modes}
	We obtain $ (w_{\textrm{KR1}},w_{\textrm{KR1}})_\sigma=(w_{\textrm{KR2}},w_{\textrm{KR2}})_\sigma=4\pi\omega I_{12},\ (w_{\textrm{KR1}},\sigma_3w_{\textrm{KR1}})_\sigma=4\pi[(\mu-n\omega)I_{12}-2gI_{14}] $, and  $ (w_{\textrm{KR2}},\sigma_3w_{\textrm{KR2}})_\sigma=4\pi[(\mu+n\omega)I_{12}-2gI_{14}] $ with $ I_{\alpha\beta}=\int_0^\infty r^\alpha f^\beta \mathrm{d}r $.  
	Therefore, the dispersion relations are given by
	\begin{align}
		\epsilon_{\textrm{KR1}}(k)=\omega+A_-k^2+O(k^4), \label{kb1disp} \\
		\epsilon_{\textrm{KR2}}(k)=\omega+A_+k^2+O(k^4), \label{kb2disp}
	\end{align}
	where
	\begin{align}
		A_{\pm}=\frac{\mu-2g\eta}{\omega}\pm n,\quad \eta:=\frac{\int_0^\infty rf^4\mathrm{d}r}{\int_0^\infty rf^2\mathrm{d}r}. \label{AKRpm}
	\end{align}
	We emphasize that $ A_{\pm} $ is always positive for any $ n $ by definition. Therefore, these modes show no Landau instability (i.e., no negative dispersion relation for positive-norm eigenstates). \\
	\indent Using Eqs.~(\ref{densityandphase})-(\ref{densityphasefluct}), the oscillation of the vortex core  and the condensate surface can be respectively estimated as
	\begin{align}
		&\bigl(r_{\text{core}}(t,z),\theta_{\text{core}}(t,z)\bigr)=\Bigl(\alpha|n|,m(\pi-kz+\epsilon t)\Bigr) \quad (\text{if } n\ne 0), \label{wkbcorepos} \\
		&r_{\text{surface}}(t,z,\theta)=r_{\text{TF}}-\alpha\cos(kz-\epsilon t+m\theta), \label{wkbsurfaceosci}
	\end{align}
	where $ \big(\epsilon,m\big)=\bigl(\epsilon_{\text{KR1}}(k),-1\bigr) $ and $ \bigl(\epsilon_{\text{KR2}}(k),+1\bigr) $ should be used for KR1 and KR2, respectively. Here, $ (r_{\text{core}},\theta_{\text{core}}) $ is obtained by solving $ |\psi+\delta\psi|=0 $ for Eq.\,(\ref{densityandphase}) with $ w_{\text{KR1,2}} $.  Also note that we define the condensate surface by the position such that the condensate density is equal to $ f(r_{\text{TF}})^2 $, where $ r_{\text{TF}} $ is the TF radius [Eq.\,(\ref{surfaceposition})]. 
Since $r_{\rm surface}(t,z,\theta +\theta_{\rm core}(t,z)) = r_{\rm TF} + \alpha \cos(\theta)$, the configuration looks static from an observer rotating with the vortex core. 
	\\
	\indent Since the KR1 and KR2 modes have the same density fluctuation $ \delta\rho $ [Eq.\,(\ref{densityphasefluct})], their difference appears in the phase fluctuation $ \delta\varphi $. It will be seen in Sec.~\ref{sec:numchk}. 

\subsubsection{$w_{\mathrm{breather}}$ --- the breather mode}
	Denoting $ I_{\alpha\beta}=\int_0^\infty r^\alpha f^\beta \mathrm{d}r $, the $\sigma$-inner products are 
	\begin{align}
		&(w_{\text{breather}},w_{\text{breather}})_\sigma=4\pi\omega^3 I_{32}, \\
		&(w_{\text{breather}},\sigma_3w_{\text{breather}})_\sigma=4\pi\left( [\mu^2+(1-n^2)\omega^2]I_{12}-2g\omega^2I_{34} \right).
	\end{align}
	Furthermore,  we can prove
	\begin{align}
		2\mu I_{12}-\omega^2I_{32}-2g I_{14}&=0, \\
		2(1-n^2)I_{12}+3\mu I_{32}-\omega^2I_{52}-4g I_{34}&=0,
	\end{align}
	by integrating  $ r^2f' \times [\text{Eq.\,(\ref{fGP})}] $ and  $ r^5(\frac{f}{r})'\times [\text{Eq.\,(\ref{fGP})}] $. Such identities are also found by following Derrick's argument of the scaling transformation for the energy functional \cite{1964JMP.....5.1252D}. \\
	\indent Thus, the dispersion relation is given by
	\begin{align}
		&\epsilon_{\text{breather}}(k)=2\omega+B k^2+O(k^4), \label{breatherdisp} \\
		&B=\frac{2\mu(3g\eta-2\mu)+\omega^4\tilde{\eta}}{4\omega(\mu-g\eta)},\quad \tilde{\eta}:=\frac{\int_0^\infty r^5f^2\mathrm{d}r}{\int_0^\infty rf^2\mathrm{d}r}, \label{breatherdisp2}
	\end{align}
	where $ \eta $ is defined by Eq.\,(\ref{AKRpm}). \\
	\indent Solving $ |\psi+\delta\psi|=0 $ for Eq.\,(\ref{densityandphase}) with $ w_{\text{breather}} $ with assuming $ f \propto r^{|n|} $ near the origin, we can check that this mode shows no vortex-core oscillation up to $ O(\alpha) $.  The surface oscillation is estimated as
	\begin{align}
		r_{\text{surface}}(t,z,\theta)\simeq r_{\text{TF}}\left[ 1-2\alpha\omega\cos(kz-\epsilon t) \right],
	\end{align}
	with $ \epsilon=\epsilon_{\text{breather}}(k) $. This expression is consistent with the  feature of Fig.~\ref{fig:intro} (B).  Here, we have assumed $ r_{\text{TF}} \gg \frac{f(r_{\text{TF}})}{f'(r_{\text{TF}})} $. \\
	\indent The node point of breathing motion, where the density oscillation vanishes, can be found by solving $ \delta\rho(r)=0 $. 
	Within the TF approximation (\ref{thomasfermi}), it is estimated as
	\begin{align}
		r_{\text{node}}\simeq\frac{\sqrt{2\mu}}{\omega}=\frac{1}{\sqrt{2}}r_{\text{TF}}. \label{breathernode}
	\end{align}
	The phase oscillation $ \delta\varphi(r) $ also vanishes at the same point. The validity of Eq.\,(\ref{breathernode}) will be checked in Figs.~\ref{fig:n0breather} and \ref{fig:n1breather}.
		
\subsubsection{$w_{\mathrm{phonon}}$ --- Bogoliubov sound wave}\label{subsec:bogoliubovsound}
	This mode yields the famous Bogoliubov phonon (or sound wave) with the linear dispersion relation $ \epsilon \propto k $ \cite{Bogoliubov}. Since  $ w_{\text{phonon}} $ is a zero-norm eigenstate, the perturbation theory needs a slight modification \cite{Takahashi2015101,PhysRevD.91.025018}, corresponding to the case of the Jordan block of size 2. The perturbation theory for general larger Jordan blocks is given in Appendix F of Ref.~\cite{Takahashi2015101}. \\
	\indent $ f(r) $ can be regarded as a function of parameters $ (\mu,\omega) $. Differentiating Eq.\,(\ref{fGP}) by $ \mu $, we find 
	\begin{align}
		\hat{L}_{0}z_{\text{phonon}}=w_{\text{phonon}},\quad z_{\text{phonon}}:=\begin{pmatrix} f_\mu \\ f_\mu \end{pmatrix}.
	\end{align}
	$ z_{\text{phonon}} $ corresponds to the generalized eigenvector of the Jordan block. Using it, we solve the finite-$ k $ Bogoliubov equation (\ref{statBogoliubov}) by a perturbation expansion $ (\hat{L}_0+\sigma_3 k^2)(w_{\text{phonon}}+\epsilon_1 k z_{\text{phonon}}+\beta k^2 w_2+\dotsb)=(\epsilon_1 k+\epsilon_2 k^2+\dotsb)(w_{\text{phonon}}+\epsilon_1 k z_{\text{phonon}}+\beta k^2 w_2+\dotsb) $. See also Ref.~\cite{Takahashi2015101}, Sec.~6.1. Then we find 
	\begin{align}
		\epsilon_1^2=\frac{(w_{\text{phonon}},\sigma_3w_{\text{phonon}})_\sigma}{(w_{\text{phonon}},z_{\text{phonon}})_\sigma}=\frac{2\int_0^\infty rf^2 \mathrm{d}r}{\partial_\mu [\int_0^\infty rf^2 \mathrm{d}r]}. \label{bogophonondisp0}
	\end{align}
	We thus obtain the linear dispersion relation
	\begin{align}
		\epsilon_{\text{phonon}}(k)=\epsilon_1 k+O(k^3). \label{bogophonondisp}
	\end{align}
	In the trap-free limit $ \omega \to 0 $, the condensate behaves as $ f^2(r\to\infty)=\frac{\mu}{2g}, $ and hence we get $ \epsilon_1 =\sqrt{2\mu} $, which is consistent with the phonon dispersion relation for a uniform condensate. \\
	\indent Here we remark on the density fluctuation of the Bogoliubov phonon. Recall Eq.\,(\ref{densityphasefluct}) again. 
	The eigenstate $ w_{\text{phonon}} $ has only a phase fluctuation $ \delta\varphi $. However, it does not mean that the Bogoliubov phonon induces no density fluctuation. It appears from the first order in $ k $ (or $\epsilon$). Using the first-order wavefunction $ w_{\text{phonon}}+\epsilon_1 k z_{\text{phonon}}+O(k^2) $,  we find $ \delta\varphi=1 $ and $ \delta\rho=\epsilon_1 k \partial_\mu(f^2) $. Thus, the generalized eigenvector $ z_{\text{phonon}} $ represents the density fluctuation.\\
	\indent Basically, the density fluctuation of phonon excitations appears from the first order in  $ \epsilon $. If it emerges from the zeroth order, it triggers the instability \cite{doi:10.1143/JPSJ.78.023001,PhysRevLett.105.035302}. 

\section{Numerical Check}\label{sec:numchk}
In this section, we numerically verify the analytical predictions derived in the last section. We plot the dispersion relations, their zero-mode eigenfunctions and density and phase fluctuations (\ref{densityphasefluct}). We also discuss the Kelvin mode appearing in the vortex state.
\subsection{The vortexless state ($n=0$)}
	\begin{figure}[tb]
		\begin{center}
		\includegraphics[width=0.95\linewidth]{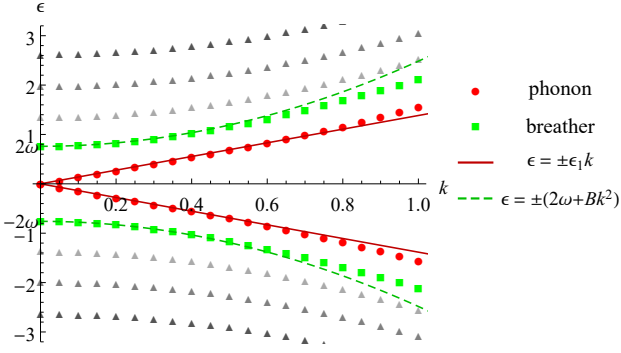}
		\caption{\label{fig:n0m0} (Color online) The dispersion relation for the eigenstates of the Bogoliubov equation with $ (n,m)=(0,0) $. We set the parameters $ g=1,\ \mu=2 $, and $ \omega=0.38 $. The red circle and green square points represent the Bogoliubov phonon and the breather excitations, respectively. The curves are given by $ \epsilon=\pm \epsilon_1 k $ and $ \epsilon=\pm(2\omega+Bk^2) $, as predicted in Eqs.~(\ref{breatherdisp}), (\ref{breatherdisp2}), (\ref{bogophonondisp0}), and (\ref{bogophonondisp}). High-energy eigenstates with no SSB origin are shown by gray triangles.  The coefficients are evaluated as $ \eta=0.624,\ \tilde{\eta}=5.81 \times 10^2,\ \epsilon_1=1.38,\ B=1.72 $. }
		\end{center}
	\end{figure}
	\begin{figure}[tb]
		\begin{center}
		\includegraphics[width=0.95\linewidth]{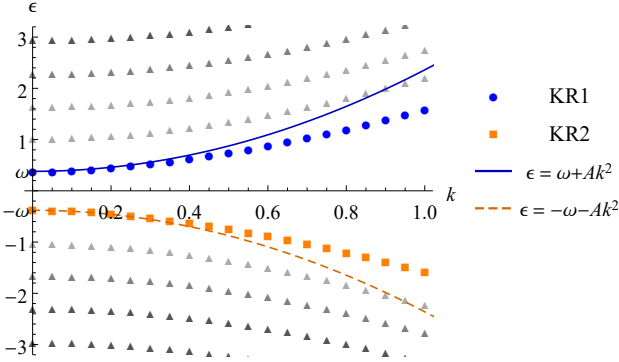}
		\caption{\label{fig:n0m1} (Color online) The dispersion relation for the eigenstates of the Bogoliubov equation with $ (n,m)=(0,-1) $. The parameters are the same as Fig.~\ref{fig:n0m0}. The blue circle and orange square points represent the KR1 and KR2 excitations. The curves represent the theoretical ones (\ref{kb1disp})-(\ref{AKRpm}), where we simply write $ A=A_+=A_- $. The numerical integration shows $ A=1.98 $. The plot for $ (n,m)=(0,1) $ is identical to this plot, but the KR1 and KR2 excitations are exchanged. }
		\end{center}
	\end{figure}

	\begin{figure}[tb]
		\begin{center}
		\includegraphics[width=0.95\linewidth]{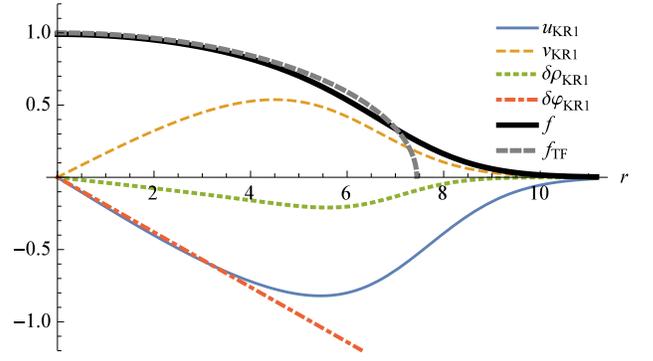}
		\caption{\label{fig:n0KR} (Color online)  Plot of the KR1 complex zero mode [Eq.\,(\ref{zerokb1})] $ w_{\text{KR1}}=(u_{\text{KR1}},v_{\text{KR1}})^T,\ \delta\rho_{\text{KR1}}=(u_{\text{KR1}}+v_{\text{KR1}})f, $ and $ \delta\varphi_{\text{KR1}}=(u_{\text{KR1}}-v_{\text{KR1}})/(2f) $ with a constant factor multiplied for visibility. We also plot the condensate wavefunction $ f(r) $ and its TF approximation $ f_{\text{TF}}(r) $ [Eq.~(\ref{thomasfermi})] for reference. The parameters are $ g=1,\ \mu=2,\ \omega=0.38 $ (the same as Fig.~\ref{fig:n0m0}), and the TF radius is  $ r_{\text{TF}}=7.44 $. There is no vortex helical motion when $ n=0 $, though it is named the ``Kelvin-ripple complex.''   Note that $ w_{\text{KR2}} $ is identical to $ w_{\text{KR1}} $.  }
		\end{center}
	\end{figure}
	\begin{figure}[tb]
		\begin{center}
		\includegraphics[width=0.95\linewidth]{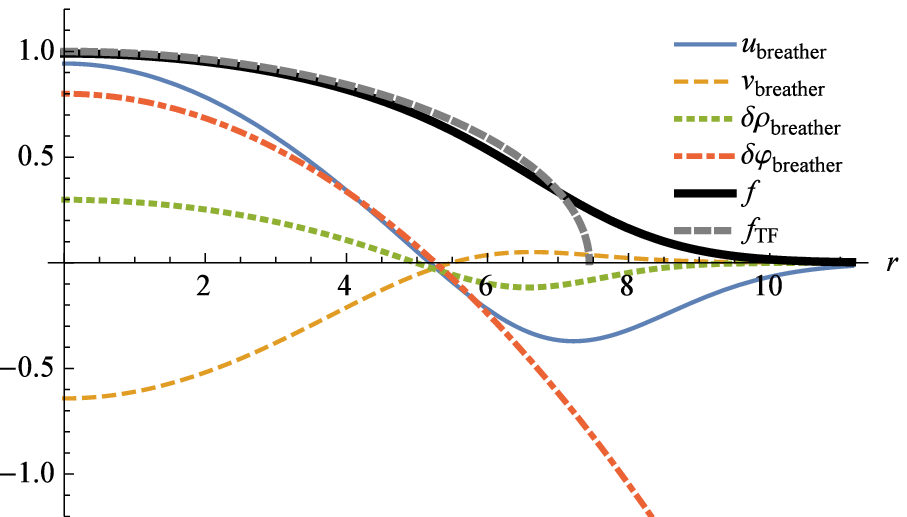}
		\caption{\label{fig:n0breather} (Color online)  Plot of the breather zero mode [Eq.\,(\ref{zerobreather})] $ w_{\text{breather}}=(u_{\text{breather}},v_{\text{breather}})^T,\ \delta\rho_{\text{breather}}=(u_{\text{breather}}+v_{\text{breather}})f, $ and $  \delta\varphi_{\text{breather}}=(u_{\text{breather}}-v_{\text{breather}})/(2f) $ with a constant factor multiplied for visibility. We also plot the condensate wavefunction $ f(r) $ and its TF approximation $ f_{\text{TF}}(r) $ [Eq.~(\ref{thomasfermi})] for reference. The same parameters are used as Fig.~\ref{fig:n0KR}. The node of the breathing motion (\ref{breathernode}) is estimated to be $ r_{\text{node}}\simeq \frac{r_{\text{TF}}}{\sqrt{2}}=5.26 $. }
		\end{center}
	\end{figure}

	We first consider the collective excitations when the background BEC is in the ground state, i.e., the state with no vortex. $ f(r) $ is a nodeless solution of Eq.\,(\ref{fGP}) with $ n=0 $, and the Bogoliubov equation is given by Eqs.~(\ref{statBogoliubov})-(\ref{statBogoliubov4}) with $ n=0 $. The profile of $ f(r) $ is well approximated by Eq.\,(\ref{thomasfermi}) except for the vicinity of the condensate surface. \\
	\indent Figures~\ref{fig:n0m0} and \ref{fig:n0m1} show the energy spectra for the eigenstates with the angular quantum numbers  $ m=0 $ and $ -1 $, respectively. Note that if $ n=0 $, the Bogoliubov equations for $ \pm m $ are completely the same. In Fig.~\ref{fig:n0m0}, we find two modes originated from the SSB of the Sch(2) symmetry, i.e., the Bogoliubov phonon and the breather mode. Their dispersion relations are well fitted by the theoretical expressions, given by Eqs.~(\ref{breatherdisp}), (\ref{breatherdisp2}), (\ref{bogophonondisp0}), and (\ref{bogophonondisp}). In Fig.~\ref{fig:n0m1}, we find the KR1 and KR2 modes. The theoretical curve is given by Eq.\,(\ref{kb1disp}) with (\ref{AKRpm}), showing a good agreement with the numerical points for small $ k $. \\
	\indent  Figure~\ref{fig:n0KR} shows a plot of the KR1 zero mode $ w_{\text{KR1}} $ [Eq.\,(\ref{zerokb1})].  Here, we must admit that the naming  ``Kelvin-ripple complex"  is a little inappropriate for the no-vortex state ($ n=0 $), because there is no vortex-core oscillation. We however use this name for brevity. The wavefunction for the KR2 mode $ w_{\text{KR2}} $ is the same as that for the KR1, up to the angular exponential factor $ \mathrm{e}^{\pm\mathrm{i}\theta} $. \\
	\indent Figure~\ref{fig:n0breather} shows the zero mode of the breather mode $ w_{\text{breather}} $ [Eq.\,(\ref{zerobreather})]. The breather mode has a node at $ r_{\text{node}}=\frac{1}{\sqrt{2}}r_{\text{TF}} $. 
\subsection{$n=1$ (the vortex state)}
	\begin{figure}[tb]
		\begin{center}
		\includegraphics[width=0.95\linewidth]{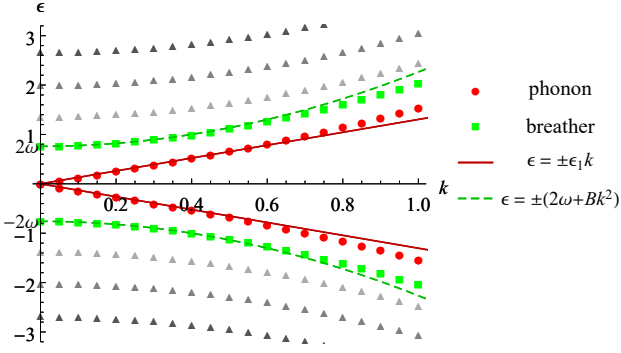}\\
		\caption{\label{fig:n1m0} (Color online) The dispersion relation for the eigenstates of the Bogoliubov equation with $ (n,m)=(1,0) $. We set the parameters  $ g=1,\ \mu=2 $, and  $ \omega=0.38 $. The red circle and green square points show the Bogoliubov phonon and breather excitations, respectively. High-energy eigenstates which have no SSB origin are shown by gray triangles. The theoretical curves are given by Eqs.~(\ref{breatherdisp}), (\ref{breatherdisp2}), (\ref{bogophonondisp0}), and (\ref{bogophonondisp}). The coefficients are evaluated as $ \eta=0.527,\ \tilde{\eta}=6.25 \times 10^2,\ \epsilon_1=1.30,\ B=1.51 $. }
		\end{center}
	\end{figure}
	\begin{figure}[tb]
		\begin{center}
		\includegraphics[width=0.95\linewidth]{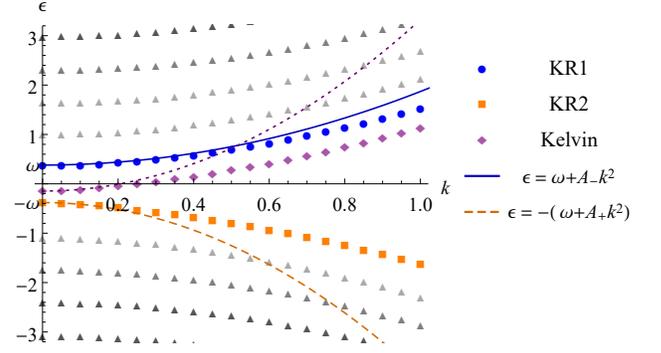}\\
		\caption{\label{fig:n1mminus1} (Color online) The dispersion relation for the eigenstates of the Bogoliubov equation with $ (n,m)=(1,-1) $. The parameters are the same as Fig.~\ref{fig:n1m0}. The blue circle and orange square points represent the KR1 and KR2 complex modes, respectively. The purple diamond points represent the Kelvin mode. The theoretical curves are given by Eqs. (\ref{kb1disp})-(\ref{AKRpm}). We numerically find $ A_-=1.49,\ A_+=3.49 $. An expression for the dispersion coefficient for the Kelvin mode is not known, but we can find it by integrating the zero-mode eigenfunction [Fig.~\ref{fig:n1kelvin}] and using the general formulas given by Eqs.~(\ref{dispersiongen}) with (\ref{dispersiongencoef}). The purple dotted curve is plotted in this way. }
		\end{center}
	\end{figure}
	\begin{figure}[tb]
		\begin{center}
		\includegraphics[width=0.95\linewidth]{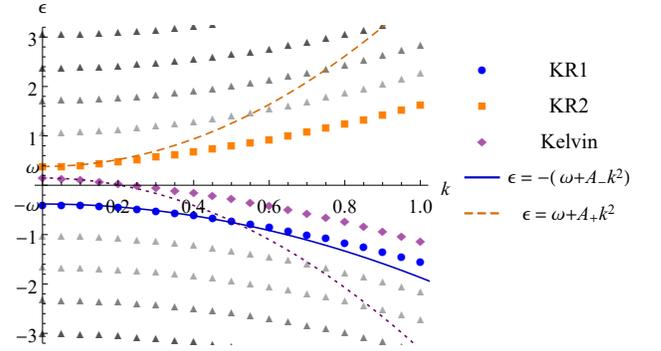}\\
		\caption{\label{fig:n1m1} (Color online) The dispersion relation for the eigenstates of the Bogoliubov equation with $ (n,m)=(1,1) $. The parameters are the same as Fig.~\ref{fig:n1m0}. This plot is the same as the vertical flip of  Fig.~\ref{fig:n1mminus1} because of the symmetry of the Bogoliubov equation.}
		\end{center}
	\end{figure}
	\begin{figure}[tb]
		\begin{center}
		\includegraphics[width=0.95\linewidth]{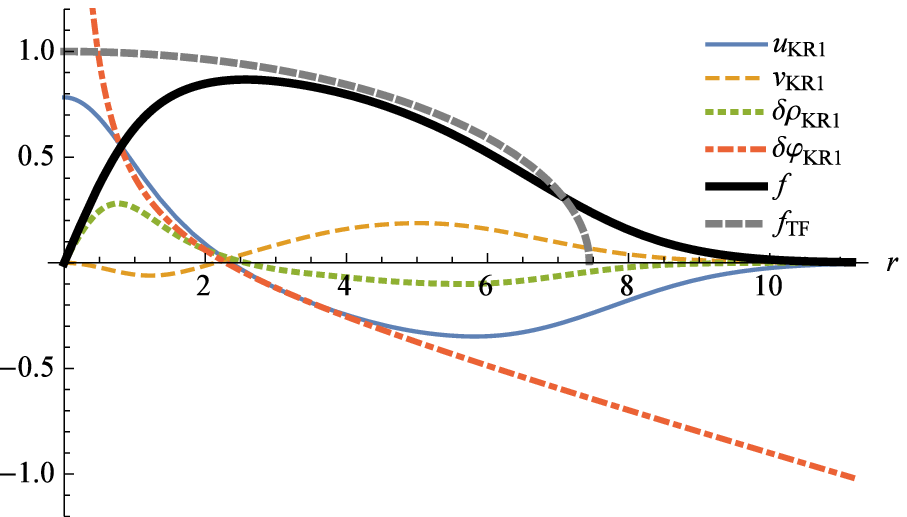}
		\caption{\label{fig:n1KR1} (Color online)  Plot of $ w_{\text{KR1}}=(u_{\text{KR1}},v_{\text{KR1}})^T $ [Eq.\,(\ref{zerokb1})], $ \delta\rho_{\text{KR1}}=(u_{\text{KR1}}+v_{\text{KR1}})f, $ and $ \delta\varphi_{\text{KR1}}=(u_{\text{KR1}}-v_{\text{KR1}})/(2f) $ with a constant factor multiplied for visibility. We also plot the condensate wavefunction $ f(r) $ and its TF approximation $ f_{\text{TF}}(r) $ [Eq.~(\ref{thomasfermi})] for reference. The parameters used are the same as Fig~\ref{fig:n1m0}. The divergent $ \delta\varphi $ around a vortex core suggests the existence of the core's helical motion. }
		\end{center}
	\end{figure}
	\begin{figure}[tb]
		\begin{center}
		\includegraphics[width=0.95\linewidth]{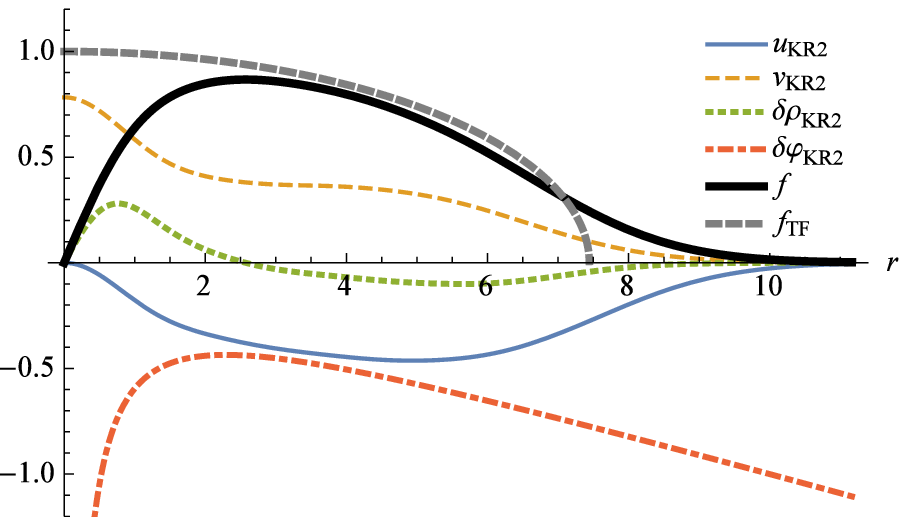}
		\caption{\label{fig:n1KR2} (Color online)  Plot of $ w_{\text{KR2}}=(u_{\text{KR2}},v_{\text{KR2}})^T $ [Eq.\,(\ref{zerokb2})],  $  \delta\rho_{\text{KR2}}=(u_{\text{KR2}}+v_{\text{KR2}})f, $ and $ \delta\varphi_{\text{KR2}}=(u_{\text{KR2}}-v_{\text{KR2}})/(2f) $  with a constant factor multiplied for visibility. We also plot the condensate wavefunction $ f(r) $ and its TF approximation $ f_{\text{TF}}(r) $ [Eq.~(\ref{thomasfermi})] for reference. The parameters used are the same as Fig~\ref{fig:n1m0}. The divergent $ \delta\varphi $ around a vortex core suggests core's helical motion. While the KR1 mode (Fig~\ref{fig:n1KR1}) has a node in the wavefunction $ (u_{\text{KR1}},v_{\text{KR1}}) $, no node is found for $ (u_{\text{KR2}},v_{\text{KR2}}) $.}
		\end{center}
	\end{figure}
	\begin{figure}[tb]
		\begin{center}
		\includegraphics[width=0.95\linewidth]{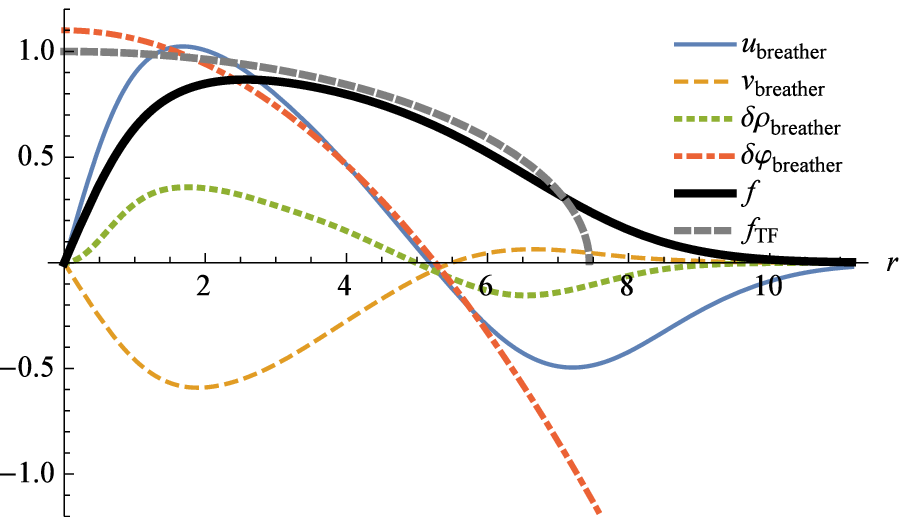}
		\caption{\label{fig:n1breather} (Color online)  Plot of $ w_{\text{breather}}=(u_{\text{breather}},v_{\text{breather}})^T $ [Eq.\,(\ref{zerobreather})],  $  \delta\rho_{\text{breather}}=(u_{\text{breather}}+v_{\text{breather}})f, $ and $  \delta\varphi_{\text{breather}}=(u_{\text{breather}}-v_{\text{breather}})/(2f) $ with a constant factor multiplied for visibility. We also plot the condensate wavefunction $ f(r) $ and its TF approximation $ f_{\text{TF}}(r) $ [Eq.~(\ref{thomasfermi})] for reference. The parameters used are the same as Fig~\ref{fig:n1m0}.  The node position (\ref{breathernode}) is given by $ r_{\text{node}}=5.26 $. The finite $ \delta\varphi $ at $ r=0 $ suggests that this mode does not show a helical motion of the vortex core. }
		\end{center}
	\end{figure}
	\begin{figure}[tb]
		\begin{center}
		\includegraphics[width=0.95\linewidth]{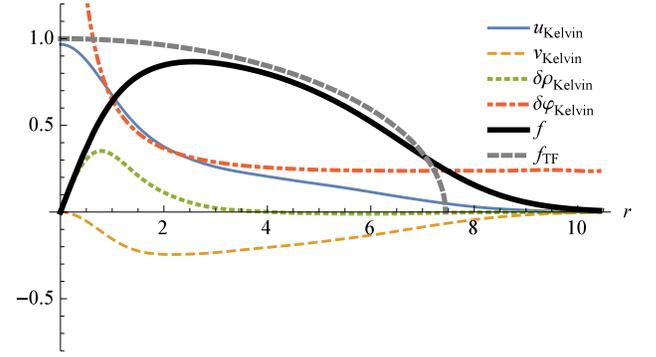}
		\caption{\label{fig:n1kelvin} (Color online)  Plot of the zero-mode wavefunction of the Kelvin mode and its phase and density fluctuaitons, corresponding to the eigenfunction for the purple diamond point of the $ k=0 $ eigenstate in Fig.~\ref{fig:n1mminus1}.  The referential condensate wavefunctions $ f(r) $ and $ f_{\text{TF}}(r) $ are also plotted. This Kelvin-mode wavefunction does not have an analytical expression, in contrast to other modes.}
		\end{center}
	\end{figure}
	We next consider the case $ n=1 $, where there exists a vortex. $ f(r) $ vanishes at the origin $ r=0 $, but the profile of the outer condensate is still well approximated by the TF wavefunction (\ref{thomasfermi}). 
	The dispersion relations for $ m=0, \pm1 $ are shown in Figs.~\ref{fig:n1m0}-\ref{fig:n1m1}, and the zero-mode eigenfunctions are plotted in Figs.~\ref{fig:n1KR1}-\ref{fig:n1kelvin}. \\
	\indent The feature of excited states for $ (n,m)=(1,0) $ shown in Fig.~\ref{fig:n1m0} is almost the same as the no-vortex case $ (n,m)=(0,0) $ (Fig~\ref{fig:n0m0}); we find the Bogoliubov phonons and the breather modes. The important difference appears in Fig~\ref{fig:n1mminus1}. From Fig.~\ref{fig:n1mminus1}, we first find that the dispersion curves of the KR1 and KR2 complex modes are not symmetric because $ A_+ \ne A_- $. Furthermore, we find a low-energy excitation shown by purple diamond dots in Fig.~\ref{fig:n1mminus1}, which cannot be predicted from the Sch(2) symmetry. As shown in Fig.~\ref{fig:n1kelvin}, the physical interpretation of this mode is just the Kelvin mode.
	This mode shows the so-called Landau instability, i.e., a positive-norm eigenstate has a negative eigenvalue. Such a character is also found for the Kelvin modes confined in the cylindrical trap \cite{Kobayashi01022014,PhysRevB.91.184501}. 
	Since in the trap-free limit ($\omega \to 0$) the Kelvin mode becomes a ``genuine'' NG mode associated with spontaneously broken translational symmetries in the presence of a vortex, the exact zero-mode eigenfunction can be found in this limit \cite{Takahashi2015101}. However, for finite $ \omega $, it does not have an exact expression, in contrast to the other modes whose zero-mode eigenfunctions are found for finite $ \omega $ using  Sch(2) symmetry.  Even if we do not know the exact dispersion relation, we can find a $ k^2 $ coefficient by the general formula (\ref{dispersiongen}) with (\ref{dispersiongencoef}). This is shown by a dotted line in Fig~\ref{fig:n1mminus1}.
	The eigenstates for  $ (n,m)=(1,1) $ in Fig.~\ref{fig:n1m1} are just a vertical flip of Fig.~\ref{fig:n1mminus1} since the eigenstate with $ (\epsilon,k,m) $ and $ (-\epsilon^*,-k,-m) $ always emerges in a pair in the Bogoliubov equation.\\
	\indent Figures~\ref{fig:n1KR1} and \ref{fig:n1KR2} show plots of the zero-mode wavefunctions (\ref{zerokb1}) and (\ref{zerokb2}) for the KR1 and KR2 complex modes, respectively. In the vortex state $ n=1 $, these two modes are distinguished by the existence or absence of the node in the wavefunctions $ (u,v) $. The divergence of the phase fluctuations $ \delta\varphi $ in the KR1 and KR2 modes indicates that the vortex core is oscillating in these modes, consistent with the expression (\ref{wkbcorepos}). Since the outer condensate is also oscillating [Eq.\,(\ref{wkbsurfaceosci})], the picture in Fig.~\ref{fig:intro} (C) is indeed realized in these modes.\\
	\indent Figure~\ref{fig:n1breather} shows a plot of the zero mode of the breather given in Eq.\,(\ref{zerobreather}). We can see that the position of the node $ r_{\text{node}} $ is well estimated by Eq.\,(\ref{breathernode}). In contrast to the KR1 and KR2 modes, this mode has finite phase fluctuation $ \delta\varphi $ at the vortex core $ r=0 $, which indicates that this excitation shows no vortex-core motion. Therefore, this mode is regarded as a pure breather, not coupled to the Kelvin mode. \\
	\indent Figure~\ref{fig:n1kelvin} is a zero-mode wavefunction of the Kelvin mode, i.e., the plot of the eigenfunction for the purple diamond dot of the $ k=0 $ eigenstate in Fig.~\ref{fig:n1mminus1}. The behavior of this mode near the vortex core is similar to the KR1 mode (Fig.~\ref{fig:n1KR1}), which is natural because both modes will reduce to the same zero mode  $ w_{x\text{-trans}}-\mathrm{i}w_{y\text{-trans}} $ in the trap-free limit $ \omega \to 0 $. However, this mode decays near the condensate surface, and therefore it can be identified as a pure Kelvin mode, corresponding to Fig.~\ref{fig:intro}~(A).

\section{Quasi-Massive-Nambu-Goldstone modes}\label{sec:QMNG}
	The collective excitations $ w_{\text{KR1}},\ w_{\text{KR2}}, $ and $ w_{\text{breather}} $ in an elongated 3D BEC, which are found by using 2D Schr\"odinger symmetry,  can be identified as both quasi- and massive NG modes. Here we explain why this is so. (Note: $ w_{\text{phonon}} $ is an ordinary NG mode associated with the $U(1)$ SSB.) \\
	\indent First, let us recall the concept of the massive NG modes \cite{Nicolis:2012vf,Nicolis:2013sga,PhysRevLett.111.021601}.  These modes emerge in the systems with an external field term which lowers the symmetry of the Lagrangian, but could be eliminated by performing some time-dependent transformation for physical variables. The Bogoliubov-theoretical treatment of these modes is available in Ref.~\cite{Takahashi2015101}, Appendix~G. If the external field is a Noether charge $ -\mu Q $, the elimination is achieved by the transformation $ \tilde{\psi}= \mathrm{e}^{-\mathrm{i}\mu Q t}\psi $. Therefore, if the symmetry group for the system with no external field is denoted by $ G $, that with the external field can be expressed as $ G'=\mathrm{e}^{\mathrm{i}\mu Q t}G \mathrm{e}^{-\mathrm{i}\mu Q t} $.   A well-known typical example is the magnetic field term $ -BS_z $ which breaks the spin-rotation symmetry. The massive NG mode is not gapless, but its existence is still robustly ensured by symmetry, and the value of the gap is determined only from the Lie algebra of the symmetry group. \\
	\indent Next, we explain the concept of quasi-NG modes based on Ref.~\cite{PhysRevD.91.025018} with a slight generalization. Let us consider a system described by several classical fields, whose Lagrangian is written by two terms: $ \mathcal{L}=\mathcal{L}_1+\mathcal{L}_2 $. We assume that each term has a group symmetry denoted by $ G_{\mathcal{L}_1} $ and $  G_{\mathcal{L}_2} $. The symmetry of the total Lagrangian is then given by $ G_{\mathcal{L}}=G_{\mathcal{L}_1}\cap G_{\mathcal{L}_2} $.  As proved in Ref.~\cite{PhysRevD.91.025018}, if we have a family of solutions parametrized by several continuous parameters, and if all elements in this family satisfies $ \mathcal{L}_2=0 $, then we can construct a parameter-derivative quasi-zero-mode solution for all generators of $ G_{\mathcal{L}_1} $. (Some of them may be a ``genuine'' zero mode originating from the true Lagrangian symmetry $ G_{\mathcal{L}} $.) Furthermore, we can find a dispersion relation for these modes by perturbation theory. In Ref.~\cite{PhysRevD.91.025018}, the theory has been constructed for  the case of
	 $ \mathcal{L}_1=-\mathcal{V},\ \mathcal{L}_2=\mathcal{L}_{\text{time}}- \mathcal{T} $, 
%	\begin{align}
%		\mathcal{L}_1=-\mathcal{V},\quad \mathcal{L}_2=\mathcal{L}_{\text{time}}- \mathcal{T}
%	\end{align}
	where $ \mathcal{L}_{\text{time}}=\int\mathrm{d}\boldsymbol{x}\sum_i\frac{\mathrm{i}(\psi_i^*\dot{\psi}_i-\dot{\psi}_i^*\psi_i)}{2} $ is a time-derivative term of the Lagrangian appearing for the Schr\"odinger-type equations, and $ \mathcal{T} $ and $ \mathcal{V} $ are kinetic  and potential terms, respectively.  $ \mathcal{H}=\mathcal{T}+\mathcal{V} $ can be regarded as a Hamiltonian. In particular, an example of the complex $ O(N) $ model, where $ G_{\mathcal{L}_1}=G_{\mathcal{V}}=O(N,\mathbb{C}) $ and $ G_{\mathcal{L}_2}=G_{\mathcal{T}}=U(3) $, has been discussed. In the condensed-matter example of the spin-2 Bose condensate \cite{Kawaguchi:2012ii}, we have $ \mathcal{L}_1=\mathcal{L}_{\text{time}}-\mathcal{H}_1 $ and $ \mathcal{L}_2=-\mathcal{H}_2 $, with 
	 $ \mathcal{L}_{\text{time}}=\int\mathrm{d}\boldsymbol{x}\sum_{m=-2}^2\frac{\mathrm{i}(\psi_m^*\dot{\psi}_m-\dot{\psi}_m^*\psi_m)}{2},\ \mathcal{H}_1=\int\mathrm{d}\boldsymbol{x} \left(\sum_{m=-2}^2 |\nabla\psi_m|^2 +c_0 \rho^2+c_1 |\Theta|^2\right),  $ and $ \mathcal{H}_2=\int \mathrm{d}\boldsymbol{x} c_2 \boldsymbol{M}^2 $, 
%	\begin{align}
%		\mathcal{L}_{\text{time}}&=\int\mathrm{d}\boldsymbol{x}\sum_{m=-2}^2\frac{\mathrm{i}(\psi_m^*\dot{\psi}_m-\dot{\psi}_m^*\psi_m)}{2}, \\
%		\mathcal{H}_1&=\int\mathrm{d}\boldsymbol{x} \left(\sum_{m=-2}^2 |\nabla\psi_m|^2 +c_0 \rho^2+c_1 |\Theta|^2\right), \\
%		\mathcal{H}_2&=\int \mathrm{d}\boldsymbol{x} c_2 \boldsymbol{M}^2,
%	\end{align}
	where $ \rho=\sum_{m=-2}^2 |\psi_m|^2 $ is a particle number density,  $ \Theta=\sum_{m=-2}^2(-1)^2\psi_m\psi_{-m} $ is a singlet pair amplitude, and  $ \boldsymbol{M}=(M_x,M_y,M_z) $ is a magnetization vector. The symmetry group for each term is given by $ G_{\mathcal{L}_1}=U(1)\times SO(5) $ and  $ G_{\mathcal{L}_2}=U(1)\times SO(3) $.  While the symmetry of the Lagrangian is $ G_\mathcal{L}=U(1)\times SO(3) $, by an appropriate choice of coupling constants $ c_0,c_1,c_2 $, we obtain the nematic phase where all the states with vanishing magnetization can be a ground state \cite{PhysRevA.61.033607,PhysRevLett.84.1066,Kawaguchi:2012ii}. The large degeneracy of this phase cannot be resolved unless the quantum many-body effects are included \cite{PhysRevLett.98.160408,PhysRevLett.98.190404,PhysRevA.81.063632}, and thus we get quasi-NG modes originating from the SSB of $ G_{\mathcal{L}_1} $, i.e., the $SO(5)$ symmetry \cite{PhysRevLett.105.230406}.\\
	\indent On the basis of the above-mentioned concepts of quasi- and massive NG modes, we can now explain why the modes $ w_{\text{KR1}},w_{\text{KR2}}, $ and $ w_{\text{breather}} $ in the harmonic trap are quasi-massive-NG modes. In the present system, we can decompose the Lagrangian into the two terms $ \mathcal{L}_1=\mathcal{L}_{\text{time}}-\mathcal{H}_1 $ and $ \mathcal{L}_2=-\mathcal{H}_2 $ with
	\begin{align}
		\mathcal{L}_{\text{time}}&=\int\mathrm{d}\boldsymbol{x}\frac{\mathrm{i}(\psi^*\dot{\psi}-\dot{\psi}^*\psi)}{2}, \\
		\mathcal{H}_1&=\int \mathrm{d}\boldsymbol{x} \left(|\partial_x\psi|^2+|\partial_y\psi|^2 +g|\psi|^4+ \frac{\omega^2(x^2+y^2)}{4}|\psi|^2\right), \\
		\mathcal{H}_2&=\int \mathrm{d}\boldsymbol{x} |\partial_z\psi|^2.
	\end{align}
	Then, as explained in Subsec.~\ref{subsec:2dSchsymGP},  $ \mathcal{L}_1 $ has a 2D Schr\"odinger symmetry modified by $ \mathcal{T}_{\text{trap}} $, that is, $ G_{\mathcal{L}_1}=\mathcal{T}_{\text{trap}}\mathrm{Sch}(2)\mathcal{T}_{\text{trap}}^{-1} $. (Recall that  $ \mathcal{T}_{\text{trap}}^{-1}=\mathcal{T}_{\text{release}} $.) Moreover, if we consider solutions with the $ z $-translational symmetry, we get $ \mathcal{H}_2=0 $, and hence, we can obtain parameter-derivative zero-mode solutions of the Bogoliubov equation originating from the SSB of $ G_{\mathcal{L}_1} $, as derived  in Secs.~\ref{sec:ssbzero} and \ref{sec:kelvin3D}. These can be regarded as quasi-NG modes in the sense that they have an origin in the SSB of the partial Lagrangian $ \mathcal{L}_1 $, but not the total $ \mathcal{L} $. The quasi-NG modes with the same origin is also found for a Skyrmion line in Ref.~\cite{PhysRevD.90.025010}. Furthermore, in the $ \mathcal{L}_1 $, the external harmonic-trap term $ \frac{\omega^2(x^2+y^2)}{4} $, which breaks the translational symmetry and hence lowers the symmetry of the total Lagrangian, can be eliminated by the transformation $ \tilde{\psi}= \mathcal{T}_{\text{release}}[\psi] $, and the zero-mode solutions are expressed as Eq.\,(\ref{zeromodew}), which are analogous to Eq.\,(G.8) in the Appendix G of Ref.~\cite{Takahashi2015101}. They have finite gaps determined by symmetry consideration, which are  $ \epsilon=\omega $ for $ w_{\text{KR1,2}} $ and $ \epsilon=2\omega $ for $ w_{\text{breather}} $. Therefore, they are also regarded as massive NG modes.  This is why we can refer to these modes as \textit{quasi-massive-Nambu-Goldstone modes}.

\section{Summary and Future Outlook}\label{sec:summary}
	To summarize, we have provided exact characteristics of the 3D collective excitations in an elongated BEC confined by a harmonic trap, using the concept of the 2D Schr\"odinger  symmetry and the Bogoliubov theory.  We found four kinds of low-energy excitations whose existence is robustly guaranteed by the Schr\"odinger-group symmetry, that is, the two KR complex modes, the one breather mode, and the Bogoliubov sound wave. We have determined their dispersion relations analytically and clarified their physical picture (Secs.~\ref{sec:2dSchsym}, \ref{sec:kelvin3D}, and \ref{sec:numchk}). We also have pointed out that the most basic excitation, i.e., the Kelvin mode, cannot be treated in terms of the SSB of the 2D Schr\"odinger symmetry.  \\
	\indent Furthermore, we have pointed out in Sec.~\ref{sec:QMNG} that the KR complex modes and the breather mode can be regarded as the quasi-massive-NG modes, extending the generalized concepts of the NG modes. \\
	\indent We have constructed the theory of the excitations propagating along the $ z $-axis in this paper. If the system length is finite in the $ z $ direction, these excitations will be observed as a standing wave, which will be a more natural setting in ultracold atomic experiments. Our formalism should be extended to the case in which the dependence of the system in the $z$-direction is small.\\
	\indent The concept of the quasi-massive-NG modes proposed in this paper should be further investigated from various viewpoints in closely related recent popular physical topics, e.g., application to the quantum turbulence \cite{PhysRevA.91.053620,2017arXiv170402566T}, the Higgs modes \cite{PhysRevD.92.045028,PhysRevA.92.043610,PhysRevD.92.055004,2016NatCo...710294Z}, and the Nambu sum rules \cite{PhysRevD.87.075016,PhysRevB.95.094515} in single and/or multi- component bosonic/fermionic 
	superfluids, and so on.  
	The splitting instability of multiple vortices in an elongated trap which may undergo the quantum turbulence was discussed in Ref.\,\cite{2015arXiv150500616T}. 
	The analytical study of the Kelvin mode for the small $g$ regime will be also worth investigating as was done in 2D in Ref.\,\cite{Biasi:2017pdp}. 
	In this paper, we have ignored the quantum many-body effects. If the quantum correction is added, the Schr\"odinger symmetry of the 2D GP system will disppear due to the quantum anomaly. Therefore, studying these effects through the dispersion relations of the breathing modes will be an important future work. 
	Another direction for future work will be a singular perturbation analysis \cite{PhysRevE.54.376} for the extrapolation of the solutions to the trap-free system $ \omega \to 0 $, where we expect only two NG modes, i.e. the Bogoliubov sound wave with linear dispersion and the Kelvin mode with logarithmic dispersion.

\section*{Acknowledgments}
This work is supported by the Ministry of Education,Culture, Sports, Science (MEXT)-Supported Program for the Strategic Research Foundation at Private Universities `Topological Science' (Grant No.\ S1511006).
The work of M.~N.~is also supported in part by the Japan Society for the Promotion of Science (JSPS) Grant-in-Aid for Scientific Research (KAKENHI Grant No.~16H03984), and a Grant-in-Aid for Scientific Research on Innovative Areas ``Topological Materials Science'' (KAKENHI Grant No.~15H05855) from the MEXT of Japan.

%%%%%%%%%%%%%%%%%%%%%%%%%%%%%%%%%%%%%%%%%%%%%%%%%%%%%%%%%%%%%%%%%%%%%%%%%%%%%%%%%%%%%%%%%%%%%%%%%%%%%%%%%%%%%%%%%%%%%%%%%%
%\bibliography{Qsch2andkelvin}% Produces the bibliography via BibTeX.

%%%%%%%%%%%%%%%%%%%%%%%%%%%%%%%%%%%%%%%%%%%%%%%%%%%%%%%%%%%%%%%%%%%%%%%%%%%%%%%%%%%%%%%%%%%%%%%%%%%%%%%%%%%%%%%%%%%%%%%%%%

\end{document}